# A review of the characteristics of 108 author-level bibliometric indicators

**Lorna Wildgaard[a]*, Jesper W. Schneider[b], Birger Larsen[c]**

[a] *Royal School of Library and Information Science, Birketinget 6, 2300 Copenhagen, Denmark*
[b] *Danish Centre for Studies in Research and Research Policy, Department of Political Science and Government, Aarhus University, Bartholins Allé 7, 8000 Aarhus C, Denmark*
[c] *Aalborg University Copenhagen, A. C. Meyers Vænge 15, 2450 Copenhagen SV, Denmark*

*Corresponding author.
*Email Addresses:* pnm664@iva.ku.dk (L. Wildgaard), jws@cfa.au.dk (J. W. Schneider), birger@hum.aau.dk (B. Larsen)

## Abstract

An increasing demand for bibliometric assessment of individuals has led to a growth of new bibliometric indicators as well as new variants or combinations of established ones. The aim of this review is to contribute with objective facts about the usefulness of bibliometric indicators of the effects of publication activity at the individual level. This paper reviews 108 indicators that can potentially be used to measure performance on individual author-level, and examines the complexity of their calculations in relation to what they are supposed to reflect and ease of end-user application. As such we provide a schematic overview of author-level indicators, where the indicators are broadly categorised into indicators of publication count, indicators that qualify output (on the level of the researcher and journal), indicators of the effect of output (effect as citations, citations normalized to field or the researcher's body of work), indicators that rank the individual's work and indicators of impact over time. Supported by an extensive appendix we present how the indicators are computed, the complexity of the mathematical calculation and demands to data-collection, their advantages and limitations as well as references to surrounding discussion in the bibliometric community. The Appendix supporting this study is available online as supplementary material.

### Keywords

Author-level Bibliometrics; Research evaluation; Impact factors; Self-assessment; Researcher performance; Indicators; Curriculum Vitaes

## Introduction

According to Whitley (2000), science operates on an "economy of reputation". Regardless of how scientists and scholars approach their *métier*, they are expected to "cultivate a reputation" and during their career they will successively be assessed individually by committees, e.g. when applying for positions and funding or are nominated for prizes and awards. The pivotal source documenting the accrual of reputation is the curriculum vitae (CV) and perhaps the single most important element in the CV is the section on research publications and thus the researcher's authorship claims. A researcher's reputational status or "symbolic capital" is to a large extent derived from his or her "publication performance". Assessing publication performance is often condensed and summarized by use of a few supposedly "objective" indicators. Especially in the last decade or so, the use of indicators at the individual author-level, for example in CVs, seems to have exploded despite previous warnings from the scientometric community (e.g., Lawrence 2003; 2008; Hirsch 2005). Essentially, there is "individual bibliometrics" before and after the introduction of the Hirsch-index, *h*. After Hirsch (2005), for a time caveats of individual bibliometrics were forgotten and the scientometric community threw themselves into indicator construction especially at the individual level. Recently, the community has returned to a more reflexive discourse where ethical aspects of individual bibliometrics as well as best practices are on the agenda (cf. plenary sessions at the ISSI 2013 and STI 2013 conferences, as well as the topic of one work task in the European ACUMEN research project[1]). In practice, administrators, evaluators and researchers seem to use indicators as never before. Administrators and evaluators for assessment purposes, whereas researchers may add indicators to their CV as a competitive move, in an attempt to show visibility in the academic community as well as the effects of publications (note, for simplicity we use the term end-user in this article to define a non-bibliometrician, who as a consumer of bibliometrics applies indicators to his or her CV).

Today public access to (not always reliable) individual-level indicators such as the *h index* variants is easy through vendors such as Google Scholar or Scopus. Alternatively, such indicators are increasingly being calculated by "amateurs" (i.e., non-bibliometricians, administrators or researchers) bibliometricians" using popular tools like *Publish or Perish*[2]. All too often, unfortunately only one indicator is provided and that is usually the most "(in)famous" ones such as the Journal Impact Factor or the *h index*. These are easily accessible and perhaps the only ones many researchers are aware of, but there are many more. Currently, we can count more than one hundred indicators potentially applicable at the individual author-level. The number of indicators seems high given the fact that it is the same few variables that are manipulated though with different algebra and arithmetic. With so many potential indicators and such widespread use, it is important to examine the characteristics of these author-level indicators in order to qualify their use by administrators and evaluators but also researchers themselves. The basic aims of the present article are to draw attention to the use of multiple indicators which allow users to tell more nuanced stories and at the same time provide a "one stop shop" where end-users can easily learn about the full range of options.

With these aims, it is imperative to examine and compare author-level indicators in relation to what they are supposed to reflect and especially their specific limitations. The usefulness of indicators has been widely discussed through the years. Common themes are disciplinary appropriateness (Batista et al. 2006; Archambault and Larivière 2010; Costas et al. 2010a), the benefits of combining indicators (van Leeuwen et al. 2003; Retzer and Jurasinski 2009; Waltman and van Eck 2009), the construction of novel indicators versus established indicators (Antonakis and Lalive 2008; Wu 2008; Tol 2009; Schreiber et al. 2012), challenges to the validity of indicators as performance is refined

1 http://research-acumen.eu/
2 http://www.harzing.com/pop.htm

through personal and social psychology in recursive behaviour (Dahler-Larsen 2012) and the complexity of socio-epistemological parameters of citations that induces a quality factor (Cronin 1984; Nelhans 2013).

There is to some extent agreement within the scientometric community that performance can only be a proxy of impact and that performance cannot be captured by a single bibliometric indicator. However outside the bibliometric community some indicators are believed to indicate both quality and impact, such as the *h index* (Hirsch, 2005) that is commonly added to CVs. The risks of researchers using indicators that condense different aspects of scientific activity in one indicator regardless of disciplinary traits are many, and the debate of the shortcomings of author-level metrics continues (Burnhill and Tubby Hille 1994; Sandström and Sandström 2009; Bach 2011; Wagner et al. 2011; Bornmann and Werner 2012). Also, results of bibliometric assessments have been shown to contribute to both positive and negative culture changes in the publishing activities of individuals, (Hicks 2004; 2006; Moed 2008; Haslam and Laham 2009; HEFCE 2009). With this is mind there is a need for indicators to be verified as to whether or not they should be used at the author-level. Depending on the aim of the assessment, a high or low score can affect the individual's chances for receiving funds, equipment, promotion or employment (Bach 2011; HEFCE 2009; Retzer & Jurasinski 2009). As consumers of author-level bibliometrics, researchers can choose the indicators they think best document their scientific performance and will draw the attention of the evaluator to certain achievements. This of course requires knowledge of the advantages and disadvantages of the indicators but also how the many different bibliometric indicators at their disposal are calculated.

Being able to practically calculate the indicator is a major part of communicating the effect of an author's body of work (referred to a as 'portfolio' in the remainder of the article). Complex calculations limit the end-user's choice of bibliometric indicators and hence which effects can be communicated and to what degree of granularity. It is therefore vital when recommending indicators to consider the *usability* of indicators suggested for measuring publications and citations. Bibliometric indicators are based on mathematical foundations that attempt to account for the quantity of publications and the effect they have had on the surrounding community. Effect is traditionally indicated as number of citations or some function hereof. However, the bibliometric indicators proposed or in use are calculated in a large variety of ways. Some of these calculations are simple whereas others are complex and presuppose access to specialised datasets. But the building block of all indicators are paper and citation counts. In addition, some more sophisticated indicators adjust the numbers for variations between fields, number of authors, as well as age or career length. In our analysis we focus, as a novel contribution, on the complexity of the indicators and the consequences for their use by individual researchers. From this point of view we apply a model of complexity to investigate the usefulness of indicators, and to what extent complex calculations limit the usefulness of bibliometric indicators. We argue that the accuracy and completeness of the assessment is limited by the complexity of the applied indicators as a key challenge in recommending bibliometric indicators to end-users. Apart from the actual mathematical foundations, other variables affect the complexity of the calculation of the indicators. For example data access and data collection, including available time and resources, increase the complexity of calculating even simple indicators (Burnhill and Tubby Hille 1994; Ingwersen 2005). Problems with data accessibility, English language bias in citation databases and missing publication and citation data limit the usability of indicators and can directly affect the complexity of the interpretation of the indicator and as such the performance of the researcher (Bach 2011; Rousseau 2006), and the goodness of fit of the mathematical model on the bibliometric data relative to end-user profiles within their field, gender and academic position is also important (Alonso et al. 2009; Iglesias and Pecharromán 2007; Wagner et al. 2011). Author-level indicators have been met with a long string of criticisms. The aim of our article is not to passively cultivate this culture of criticism but to actively contribute with objective facts about the usefulness of bibliometric indicators of the effects of publication activity. We are aware of the many caveats but will not discuss them further in this article and focus instead on the issue of complexity. Note also that we limit our study to indicators of the effect of traditional types of publications within the academic community or public sphere, as attempting to review all types of indicators and activities, although needed, is beyond the scope of the present article. Given these aims and caveats, our research questions are:

- *Which author-level bibliometric indicators can be calculated by end-users?*
- *Is it possible to understand what the indicators express?*

The article is structured as follows, the next section provides the background for author-level indicators and the theoretical framework we apply; the subsequent section outlines the methodology of the analysis, including an outline of the analytical framework we use based on Martin and Irvine (1983), and the final two sections contain extensive presentations of and discussions of the analyses and the results.

## Methodology

We chose to limit the types of author-level indicators to indicators of the effects of publication activity, resulting in the exclusion of indicators of other important activities such as societal impact, web presence, leadership skills, technical skills, teaching activities, innovation, etc. We included famous indicators that are suggested for use, indicators that are direct adaptations of these known indicators and novel indicators that have been introduced to the bibliometric community but only tested on limited datasets. Novel indicators are included in this review as they are imaginative, attempt to correct for the shortcomings of established indicators and provide alternative ideas to assessment.

Beginning with known works on author-level assessment we identified indicators by exploring the history and development of author-level bibliometrics discussed in Directorate General for Research (2008), Schreiber (2008a), De Bellis (2009), Sandström and Sandström (2009) and Bach (2011). We used citation and reference chasing to find previously unidentified indicators. Supplementary information about the extent the indicators measure what they purport to measure were sourced using the terms (*bibliometri* OR indic*) AND (individual OR micro* OR nano**) in Thomson Reuters Web of Science® (WoS) and in the Royal School of Library and Information Science's electronic collection of information science journals. Technical papers that analyse the properties of groups of indicators in cluster or factor analyses proved particularly useful. Google Scholar was searched to retrieve for instance national papers, reports, book chapters and other web-based material, such as websites devoted to bibliometric indicators, mediated bibliometric recommendations from ministerial reports, teaching materials and library websites.

### Categories of publication indicators

We designed a simple typology of publication and effect indicators that builds on the work of Martin and Irvine (1983). This well-known work recommended thirty years ago a simple model of counting citations and papers to evaluate success and differences in research performance. The simplicity of their model of performance assessment interprets citations as indicators of impact, not quality or importance; presents a range of indicators each focussing on different aspects of research performance and the model clearly illustrates that indicators should be applied to matched research groups, i.e. to compare like with like. We diverge from their model of indicating the performance of research groups, as we extend their model to author-level assessment. We categorize the methods of publication and citation count at the author-level as follows:

1) *Indicators of publication count (output)*: methods of counting scholarly and scientific works published or unpublished depending on the unit of assessment.
2) *Indicators that qualify output as Journal Impact*: impact of a researcher's chosen journals to suggest the potential visibility of the researcher's work in the field in which he/she is active.
3) *Indicators of the effect of output:*
    a. *Effect as citations:* methods of counting citations, whole or fractional count.
    b. *Effect of output normalized to publications and field:* Indicators that compare the researcher's citation count to expected performance in their chosen field.
    c. *Effect of output as citations normalized to publications and portfolio:* Indicators that normalize citations to the researcher's portfolio.
4) *Indicators that rank the publications in an individual portfolio:* indicators of the level and performance of all of the researcher's publications or selected top performing publications. These indicators rank publications by the amount of citations each publication has received and establish a mathematical cut-off point for what is included or excluded in the ranking. They are subdivided into the following:
    a. *h-dependent indicators*
    b. *h-independent indicators*
    c. *h adjusted to field*
    d. *h adjusted for co-authorship*
5) *Indicators of impact over time:* indicators of the extent a researcher's output continues to be used or the decline in use.
    a. *Indicators of impact over time normalized to the researcher's portfolio*
    b. *Indicators of impact over time normalized to field*

The broad categorization of indicators helps us keep the analysis simple and at the same time enables us to identify relationships between the indicators. The indicators identified in the search strategy were grouped according to the aspect of the effect of publication activity that the developers of each specific indicator claim the indicators to measure. As indicators are evolutionary and supplement each other, they cannot in practice be restricted to just one category. Accordingly we agree with Martin and Irvine (1983) that assessment of research performance can be defined in many ways and, particularly in the assessment of publications and citations of individuals, combining indicators from different categories to capture the many different facets of publication activity is recommended.

## Judgement of complexity

For each indicator we investigated its intended use, calculation and data requirements. We assume that the end-user has a complete publication list and would only need to find publication data on known documents, citations and calculate the indicator. Each retrieved paper describing the components of indicators was read and the indicators were graded on two aspects of complexity on a 5 point numerical scale namely 1) the availability of citation data and, 2) the intricacy of the mathematical model required to compile the indicator, see Table 1 below. Data requirements were simple to judge, however level of computation proved difficult as mathematical capabilities are individual. Therefore in cases of doubt we calculated the indicator to understand the mathematical foundations and reach consensus about the indicator's level of complexity. All indicators that scored $\leq 3$ were calculated to check the complexity score was defendable. As this is a subjective model of scoring complexity, we support our judgements in the extensive appendix that describes the calculations, advantages and disadvantages of each indicator (Online Resource 1). The appendix was our decision tool through-out the judgement process and is published as a supplementary file online.

**Table 1.** Five point scale used in assessing two aspects of complexity of bibliometric indicators

**EASY**

| Level | Citation Data Collection |
|---|---|
| 1 | No citation data needed |
| 2 | Individual's citations or ready-to-use journal indicators from structured sources |
| 3 | Citing articles, journal, category, field, or world citation data needed, from structured sources |
| 4 | Citation data from unstructured sources |
| 5 | Citation data not readily available |

+

| Level | Calculation of indicator |
|---|---|
| 1 | Raw count only |
| 2 | Single, simple ratio or linear model |
| 3 | Simple, multiple calculation, e.g. repeated simple linear or ratio calculations |
| 4 | Advanced multiple calculation, use of weighted parameters gamma or delta that the user must define dependent on the discipline, time interval, velocity or other corrective factors. |
| 5 | Advanced multiple calculations and transformation of data |

**DIFFICULT**

Our scoring of indicators might result in a set of indicators identified as useful which have lower granularity and sophistication. This represents a balance between, on the one hand, using indicators that are as accurate as possible and measure what they purport to measure, and on the other recommending indicators that not so complex as to deter end-users to use them in practice. The indicators have to measure what they purport to measure of course, however, usability is lost if correct measurement requires data that is not readily available to the end-user, difficult mathematical calculations, and intricate interpretations of complicated data output. We choose to categorise any indicator that scores 4 or above on either of the two complexity criteria as too complex for end-users to apply in practise – and thus not useful.

# Results

We identified 108 indicators recommended for use in individual assessment of publication activity and impact. They are presented in tables in the appendix (Online Resource 1) where we briefly describe how each indicator is calculated, provide bibliographic references and discuss what they are designed to indicate, their limitations, advantages, their complexity scores and give comments on their functionality found in related literature. Table 2 below presents an overview of the assessments of complexity, followed by Tables 3 to 13 with details about each indicator. Indicators are shown in *italics* in the text.

## Overview of the identified indicators

Out of the 108 indicators we identified as potentially applicable on the level of individual researchers, one third of the indicators are adaptions of the *h index* (35/108). In Table 2 we present the indicator category, the amount of indicators in that category, the number of indicators that scored ≤ 3 in data collection and calculation and in the final column the number of indicators that scored ≥4 in either data collection or calculation.

**Table 2.** The amount and complexity of indicators in each category

| Category | No. of Indicators | Complexity ≤ 3 | Complexity ≥4 |
|---|---|---|---|
| 1) *Publication Count* | 15 | 15 | 0 |
| 2) *Journal Impact* | 20 | 16 | 4 |
| 3a) *Effect of output as citations* | 11 | 9 | 2 |
| 3b) *Effect of output as citations normalized to publications and field* | 8 | 7 | 1 |
| 3c) *Effect of output as citations normalized to publications in the portfolio* | 6 | 6 | 0 |
| 4a) *h-dependent indicators* | 16 | 10 | 6 |
| 4b) *h-independent indicators* | 7 | 4 | 3 |
| 4c) *h adjusted to field* | 5 | 2 | 3 |
| 4d) *h adjusted for co-authorship* | 5 | 2 | 3 |
| 5a) *Impact over time normalized to portfolio* | 12 | 8 | 4 |
| 5b) *Impact over time normalized to field* | 3 | 0 | 3 |
| Total | 108 | 79 | 29 |

## Summary of complexity scores

Overall, our complexity scoring resulted in 79/108 indicators scoring ≤3 in both collection of data and calculation, and thus we judged them potentially useful for end-users. The remaining 29 indicators were scored as ≥4 in *either* effort to collect citation data or in the calculation itself. Though possibly more accurate and superior measures, these indicators require either special software, e.g. *h index sequences and matrices, hT, co-authorship network analysis*; access to sensitive data, e.g. *knowledge use;* access to restricted data, e.g. *scientific proximity, citations in patents*; no agreement on weighting factors, correcting factors or values of alpha parameters, e.g. *hα, gα, a(t), prediction of article impact;* or advanced multiple calculations, e.g. *hp, hap, DCI, dynamic h, rat h, rat g*. Consequently, these indicators, amongst others, are not considered applicable by an end-user.

The tables in the following analytical summary are limited to the acronym and full name of the indicator; a short description of what it is designed to indicate as defined by the inventor of the indicator, supported with a bibliographic reference and the results of the complexity analysis where **Col** indicates complexity of data collection and **Cal** indicates complexity of data calculation. The indicators that we judged too complex to be useful are highlighted in grey. Primarily indicators that scored ≤ 3 are discussed in the text following each table; however some complex indicators are discussed in categories where no simple indicators were identified. The sections are annotated to help the reader refer back to our categories of publication indicators (see the Methodology section and Table 2 above).

## Publication Count, category 1

Fifteen indicators of publication count were identified, all with a complexity score ≤ 2, Table 3. These are simple counting or ratio models that treat contribution to a publication as equally or fractionally distributed across authors.

*P* is the raw count of publications, while $P_{isi}$, $P_{ts}$, count only publications indexed in predetermined sources, which can of course be adapted to any bibliographical database, specific selection of journals or publishers of books. Likewise *weighted publication type* and *patent applications* also account for types of publication judged locally important, showcase specific skills of the researcher or focus on publications deemed as a higher scientific quality relative to the specialty of the researcher. *Dissemination in the public sphere* counts publication and dissemination activities via other channels than academic books or articles. This indicator of publication count is just one of the indicators suggested by Mostert et al. (2010) in their questionnaire tool to measure societal relevance which also includes standardised weighting schemes to accommodate certain activities in the field the researcher is active in. All the aforementioned counting methods assume an equal distribution of contribution across all authors of a publication. The following indicators share the credit for a *publication fractionally* (equal credit allotted to all co-authors), *proportionally* (credit is adjusted to author position on the byline), *geometrically* (twice as much credit is allotted to the *i*th author as to the (*i* + 1)th author) or *harmonically* (credit is allocated according to authorship rank in the byline of an article and the number of coauthors). *Noblesse oblige* and *FA* prioritize the last and first author in crediting a publication. Correct factional counting should support level of collaboration, not just an integer number symbolizing a share but of course this increases the complexity of the indicator, as data collection would also have to include author declarations. *Co-author* and *co-publication* counts can be extended into analyses of collaboration, networks or even cognitive orientation that identify the frequency a scientist publishes in various fields and if combined with a similar citation study, their visibility and usage. These are, however, outside the scope of this review.

**Table 3.** Indicators of publication count.

| Publication Count (1) | Designed to indicate | Complexity | |
|---|---|---|---|
| | | Col | Cal |
| **P** (total publications) | Count of production used in formal communication. | 1 | 1 |
| **FA** (first author counting) | Credit given to first author only. | 1 | 1 |
| **weighted publication count** | A reliable distinction between different document types. | 1 | 1 |
| **patent applications** (Okubu 1997) | Innovation. | 1 | 1 |
| **Dissemination in public sphere** (Mostert et al. 2010) | Publications other than scientific & scholarly papers. | 1 | 1 |
| **co-publications** | Collaboration on departmental, institutional, international or national level & identify networks. | 1 | 1 |
| **co-authors** | Indicates cooperation and growth of cooperation at inter-and national level. | 1 | 1 |
| **P (publications in selected databases) e.g. Pisi,** | Publications indexed in specific databases, output can be compared to world subfield average. | 1 | 2 |
| $\mathbf{P_{ts}}$ (publications in selected sources) | Number of publications in selected sources defined important by the researcher's affiliated institution. | 1 | 2 |
| **fractional counting on papers** | Shared authorship of papers giving less weight to collaborative works than non-collaborative ones. | 1 | 2 |
| **proportional or arithmetic counting** | Shared authorship of papers, weighting contribution of first author highest and last lowest. | 1 | 2 |
| **geometric counting** | Assumes that the rank of authors in the by-line accurately reflects their contribution. | 1 | 2 |
| **harmonic counting** | The 1$^{st}$ author gates twice as much credit as the 2$^{nd}$, who gets 1.5 more credit than the 3$^{rd}$, who gets 1.33 more than the 4$^{th}$ etc. | 1 | 2 |
| **noblesse oblige** (last author count) | Indicates the importance of the last author for the project behind the paper. | 1 | 2 |
| **cognitive orientation** | Identifies how frequently a scientist publishes (or is cited) in various fields; indicates visibility/use in the main subfields and peripheral fields. | 2 | 1 |

## Qualifying output as Journal Impact, category 2

Even though journal impact indicators were originally designed as measures of journal or group impact, we have found in the literature that they are applied at an author-level to suggest the visibility of a researcher's work, Table 4. We are aware that many more impact factors are available, and that these are analyzed in detail elsewhere (e.g. Haustein 2012). We therefore only include the main types. Publications in selected journals,, $P_{tj,.}$is the only journal impact factor *designed* for use at the author-level; $P_{tj}$ has the advantage that it is entirely independent of subject categories in WoS. It is calculated using journals identified as important for the researcher's field or affiliated institution by the department or university. The journal Impact factors *JIF*, *AII*, *CHL* and *ACHL* are easily available to the end-user through WoS Journal Citation Reports (JCR). *JIF* is the average citation per article, note or review published by the journal over the previous two years calculated using Thompson Reuter's citation data. At the author-level it is commonly used to measure the impact factor of the journals in which a particular person has published articles. *NJP* ranks journals by *JIF* in a JCR subject category. If a journal belongs to more than one category, an average ranking is calculated. The lower the *NJP* for a journal, the higher its impact in the field. Similar to *NJP* is *IFmed*, which is the median value of all journal Impact Factors in the JCR subject category. However, unlike *IFmed*, *NJP* allows for inter-field comparisons as it is a field normalized indicator (Costas et al. 2010a). Misuse in evaluating individuals can occur as there is a wide variation from article to article within a single journal. Hence, it is recommended in JCR to supplement with the *AII*, *CHL* and *ACHL* indicators which indicate how quickly the average article in the journals are cited, i.e. how quickly the researcher's papers are visible in the academic community. An alternative to *JIF* is the *DJIF*, which identifies articles published in a journal by the researcher in a certain year and the average number of citations received during the 2 or more following years. As a result, *DJIF* reflects the actual development of impact over time of a paper or set of papers. Even though the data collection is more resource demanding, the benefit for the researcher is that it can be calculated for one-off publications, such as books or conference proceedings. *SJR* and *SNIP* (source normalized impact per publication indicator) are journal impact factors based on data from Scopus instead of WoS, and as such include potentially more data on European publications. *SJR* is based on a vector space model of journals co-citation profiles to provide an indication of journal prestige and thematic relation to other journals independent of WoS subject categories. With its longer publication and citation window of three years and the normalization of citations *SNIP* attempts to correct for differences in citation practices between scientific fields.

**Table 4.** Indicators that ualify output using Journal Impact Factors.

| Journal Impact (2) | Designed to indicate | Complexity | |
|---|---|---|---|
| | | Col | Cal |
| **$P_{tj}$** (Rehn et al. 2007) | Performance of articles in journals important to (sub)field or institution. | 1 | 2 |
| **ISI JIF (SIF) synchronous IF** | Average number of citations a publication in a specific journal has received limited to WoS document types and subject fields. | 2 | 1 |
| **SNIP** (Moed 2010; Waltman et al. 2012) | Number of citations given in the present year to publications in the past three years divided by the total number of publications in the past three years normalized to field. Based on Scopus data. | 2 | 1 |
| **immediacy index** | Speed at which an average article in a journal is cited in the year it is published. | 2 | 1 |
| **AII, aggregate Immediacy Index** | How quickly articles in a subject are cited. | 2 | 1 |
| **CHL, cited half-life & ACHL, aggregate cited half-life** | A benchmark of the age of cited articles in a single journal. | 2 | 1 |
| **IFmed** (Costas et al. 2010a) | Median impact factor of publications. | 2 | 2 |
| **SJR, Scimago journal rank** | Average per article PageRank based on Scopus citation data. | 2 | 1 |
| **AI, article influence score** | Measure of average per-article citation influence of the journal. | 2 | 1 |
| **NJP, normalised journal position** (Bordons and Barrigon 1992; Costas et al. 2010a) | Compares reputation of journals across fields. | 2 | 2 |
| **DJIF, diachronous IF** (Ingwersen et al. 2001) | Reflects actual and development of impact over time of a set of papers. | 3 | 2 |
| **CPP/FCSm** (Costas et al. 2010a) | Impact of individual researchers compared to the world citation average in the subfields in which the researcher is active. | 3 | 3 |
| **CPP/JCSm** | Indicates if the individual's performance is above or below the average citation rate of the journal set. | 3 | 3 |
| **JCSM/FCSm** (Costas et al. 2009; 2010a) | Journal based worldwide average impact mean for an individual researcher compared to average citation score of the subfields. | 3 | 3 |
| **C/FCSm** (van Leeuwen et al. 2003) | Applied impact score of each article/set of articles to the mean field average in which the researcher has published. | 3 | 3 |
| **prediction of article impact** (Levitt and Thelwall 2011) | Predictor of long term citations. | 3 | 4 |
| **co-authorship network analysis** (Yan and Ding 2011) | Individual author-impact within related author community. | 2 | 5 |
| **$\bar{c}f$, item oriented field normalized citation score average** (Lundberg 2009) | Item orientated field normalised citation score. | 3 | 4 |
| **%HCP** (Costas et al. 2010a) | Percent papers in the 20% most cited in the field. | 3 | 4 |

*CPP/FCSm*, *JCSm/FCSm* are used together to evaluate individual by Costas et al. (2010a) to indicate the impact profile of individuals. The observed impact of a researcher was indicated by normalizing the *%HCP*, *CPP* and *CPP/FCSm* indicators, while the quality of the journals the individual publishes in was indicated using normalized *IFmed*, *NJP* and *JCSm/FCSm*. As citation rates are increasing and disciplines evolving it is important to normalize the measured impact of researchers to their specialty or discipline. Therefore citations to journals are calculated, as a proxy set for specialty or disciplinary averages using indicators *CPP/JCSm* or *C/FCSm*. Normalization allows for inter-field comparisons (Costas et al. 2010a)

## Effect of Output, category 3

*Effect as citations, 3a*

Nine of the 11 identified indicators counting citations were judged useful in assessment, ≤3. *C+sc*, and database dependent counting calculate the sum of all citations for the period of analysis, while *C*, *C-sc*, adjust the sum for self-citations. Self-citations, *sc*, are relatively simple to collect and calculate but definition can be problematic. *Sc* can be citations by researchers to their own work, but also citations by their co-authors or even affiliated institution. The number of not cited papers, *nnc* is used to illustrate if the citations a researcher has received come from a few highly recognized papers, a stable cited body of work or a group of papers that pull *CPP* in a negative direction. Likewise *MaxC* indicates the most highly cited paper, which can skew indicators based on citation averages but also identify the researcher's most visible paper. Another simple indicator of most visible papers is the *i10 index*, which indicates the amount of papers that have received at least 10 citations each. Just as in *fractional counting of publications*, there are methods to adjust citation count according to the amount of authors to ensure a "fair" distribution of citations, again these assume at the simplest level that authors contribute equally to the paper. Further, they have the benefit of adjusting for the effect of multi-authorship that can in some fields heavily inflate the total amount of citations a researcher receives.

**Table 5.** Indicators of the effect of output as citations.

| Effect as citations (3a) | Designed to indicate | Complexity | |
|---|---|---|---|
| | | Col | Cal |
| **nnc** | Number of publications not cited. | 1 | 1 |
| **database dependent counting** (Scimago Total Cites, WOS, Scopus) | Indication of usage by stakeholders for whole period of analysis in a given citation index. | 2 | 1 |
| **C + sc** (total cites, inc. self-citations) | Indication of all usage for whole period of analysis. | 2 | 1 |
| **i10 index**, Google Scholar metric | The number of publications with at least 10 citations | 2 | 1 |
| **C** (typically citations in WOS, minus self cites) | Recognised benchmark for analyses. Indication of usage by stakeholders for whole period of analysis. | 2 | 2 |
| **Sc** | Sum of self-citations. | 2 | 2 |
| **fractional citation count** (Egghe 2008) | Fractional counting on citations removes the dependence of co-authorship. | 2 | 2 |
| **C-sc** (total cites, minus self-cites) | Measure of usage for whole period of analysis. | 2 | 2 |
| **MaxC** | Highest cited paper. | 2 | 2 |
| **citations in patents** (Okobu 1997) | Citations or use in new innovations. | 4 | 1 |
| **knowledge use** (Mostert et al. 2010) | Citations in syllabus, schoolbooks, protocols, guidelines, policies and new products. | 5 | 1 |

*Effect as citations normalized to publications and field, 3b*

Identifying the top publications in a field requires the user to design field benchmarks, which is time consuming, or alternatively accept ready-to-use standard field indicators. These standard indicators are based on subject categories in citation indices that may not represent the specialty or nationality of the researcher. Ratio-based indicators account for the amount of citations relative to publications to a fixed field value, *Field Top %, E(Ptop), A/E(Ptop), Ptop*.

**Table 6.** Indicators of the effect of output as citations normalized to publications and field.

| Effect as citations normalized to publications and field (3b) | Designed to indicate | Complexity | |
|---|---|---|---|
| | | Col | Cal |
| **tool to measure societal relevance** (Niederkrotenthaler et al. 2011) | Aims at evaluating the level of the effect of the publication, or at the level of its original aim. | 1 | 1 |
| **number of significant papers** | Gives idea of broad and sustained impact. | 2 | 1 |
| **Field Top %** citation reference value | World share of publications above citation threshold for n% most cited for same age, type and field. | 3 | 3 |
| **E(Ptop)** (expected % top publications) | Reference value: expected number of highly cited papers based on the number of papers published by the research unit. | 3 | 3 |
| **A/E(Ptop)** (ratio actual to expected) | Relative contribution to the top 20, 10, 5, 2 or 1% most frequently cited publications in the world relative to year, field and document type. | 3 | 3 |
| **IQP, Index of Quality and Productivity** (Antonakis and Lalive 2008) | Quality reference value; judges the global number of citations a researcher's work would receive if it were of average quality in its field. | 3 | 3 |
| **Ptop** (percent top publications) | Identify if publications are among the top 20, 10, 5, 1% most frequently cited papers in subject/subfield/world in a given publication year. | 3 | 3 |
| **Scientific proximity** (Okubu 1997) | Intensity of an industrial or technological activity. | 5 | 2 |

The 'Index of Quality and Productivity', *IQP*, corrects for academic age, calculates user defined field averages (based on the journals the researcher has published in) and calculates the ratio expected citations to actual citations. This produces indicators of the amount of papers researchers have in their portfolio that perform above the average of the field and how much more they are cited than the average paper. *Number of significant papers* is an indicator on the same theme as *IQP* and uses a field benchmark approach where the number of papers in the top 20% of the field is considered "significant"; note the caveats for using mechanical significance tests for such decisions (e.g., Schneider 2013; forthcoming). Alternatively a more qualitative approach for identifying *number of significant papers* is adjusting for seniority, field norm and publication types. However this approach can randomly favour or disfavour researchers. Niederkrotenthaler et al.'s self-assessment *tool to measure societal relevance* attempts to qualify the effect of the publication or its original aim in society by assessment knowledge gain, stakeholders and the researcher's interaction with them. The success of the indicator is dependent on the effort of the researcher to complete the application and assessment forms for the reviewer. It is debateable if this questionnaire is a "bibliometric indicator", but we include it as it attempts to quantify the level of the effect the publication or the original aim has on society by evaluating knowledge gain, awareness, stakeholders, and the researcher's interaction with them.

*Effect as citations normalized to publications in portfolio, 3c*

The average cites per paper *CPP*, percent self-citations *%SELFCIT* and percent non-cited publications, *%PNC*, are ratio-based indicators which account for the amount of citations relative to the amount of publications in the portfolio. *%PNC* is an indication of articles that have not been cited within a given time frame while *%nnc* is simply the percent papers in the portfolio that have not been cited. The indicator *Age of citations* assesses how up-to-date or "current" a publication is for the academic community by measuring the age of the citations it receives. This indicates if the citation count is due to articles written a long time ago and are no longer cited OR articles that continue to be cited.

The calculation of these indicators is simple, but it is important that the end-user states which citation index the citation count is based on, as a researcher's papers could be uncited in one database but well cited in another dependent on the indexing policy and coverage of the source.

**Table 7.** Indicators of the effect of output as citations normalized to publications in the researcher's portfolio.

| Effect as citations normalized to publications in portfolio (3c) | Designed to indicate | Complexity | |
|---|---|---|---|
| | | Col | Cal |
| **%nnc** | Percent not cited. | 1 | 2 |
| **%PNC** (percent not cited) | Share of publications never cited after certain time period, excluding self-citations. | 2 | 2 |
| **CPP** (cites per paper) | Trend of how cites evolve over time. | 2 | 2 |
| **MedianCPP** | Trend of how cites evolve over time, accounting for skewed citation pattern. | 2 | 2 |
| **Age of citations** | If a large citation count is due to articles written a long time ago and no longer cited OR articles that continue to be cited. | 3 | 2 |
| **%SELFCIT** | Share of citations to own publications. | 3 | 2 |

## Indicators that rank the publications in the researcher's portfolio, category 4

It is interesting to assess if the publications in the portfolio contain a core of high impact publications. This is done by ranking work within the portfolio by the amount of times cited to create cumulative indicators of a researcher's production and citations. The most commonly used of these is Hirsch's *h index* (Hirsch 2005) which has been corrected and developed since its creation.

*h-dependent indicators, 4a*

Ten of the sixteen *h*-dependent indicators scored ≤3 in complexity of calculation and data collection: *h*, *m*, *e*, *hmx*, *Hg*, $h^2$, *a*, *r*, *ħ*, $Q^2$. As these are dependent on the calculation of *h index*, they suffer from the same inadequacies as *h*. The advantages and disadvantages of *h* are explained in detail in i.a. (Costas and Bordons 2007; Alonso et al. 2009; Schreiber et al. 2012). *A*, *ħ*, *m* are recommended for comparison across field or seniority. The indicators have subtle differences in their adaptions of the *h index* and which sub-set of publications from a researcher's portfolio is used. *h* ranks publications in descending order to rank them. *h* is defined where the rank and number of citations are the same or higher. The publications that are ranked equal or higher than *h* are called the *h*-core and regarded as the productive articles. Roughly proportional to *h* is *ħ*, which is the square root of half of the total number of citations to all publications.

*R*, *hg*, $h^2$, *e*, $Q^2$ and *m* adjust for the effects or discounting of highly cited papers in the calculation of *h*; *e* calculates excess citations of articles in the *h*-core, *A* is the average number of citations to the *h*-core articles whereas *m* is the median number of citations; *R* is the square root of *A*, *hg* is the square root of the sum of *h* multiplied by the *g* index while $h^2$ is proportional to the cube root of citations; $Q^2$ is the square root of the sum of the geometric mean of the *h index* multiplied by the median number of citations to papers in the *h*-core. As such $Q^2$ claims to provide a balanced indication of the number and impact of papers in the *h*-core. Finally, *hmx* simply recommends the researcher refer to their *h index* scores measured across Google Scholar, WOS and Scopus on their CVs.

**Table 8.** Indicators that rank publications in the portfolio, *h*-dependent indicators.

| *h*-dependent indicators, (4a) | Designed to indicate | Complexity | |
|---|---|---|---|
| | | Col | Cal |
| **h index** (Hirsch 2005) | Cumulative achievement. | 2 | 2 |
| **m index** (Bornmann et al.2008) | Impact of papers in the *h*-core. | 2 | 2 |
| **e index** (Zhang 2009) | Complements the *h index* for the ignored excess citations. | 2 | 2 |
| **hmx index** (Sanderson 2008) | Ranking of the academics using all citation databases together. | 2 | 2 |
| **Hg index** (Alonso et al.2009) | Greater granularity in comparison between researchers with similar *h* and *g* indicators. *The g index* is explained in table 9. | 2 | 3 |
| **$h^2$ index** (Kosmulski 2006) | Weights most productive papers but requires a much higher level of citation attraction to be included in index. | 2 | 3 |
| **A index** (Jin 2006; Rousseau 2006) | Describes magnitude of each researcher's hits, where a large a-index implies that some papers have received a large number of citations compared to the rest, (Schreiber, Malesios, and Psarakis, 2012) . | 2 | 3 |
| **R index** (Jin et al. 2007) | Citation intensity and improves sensitivity and differentiability of *A* index. | 2 | 3 |
| **ħ index** (Miller 2006) | Comprehensive measure of the overall structure of citations to papers. | 2 | 3 |
| **$Q^2$ index** (Cabrerizoa et al. 2012) | Relates two different dimensions in a researcher's productive core: the number and impact of papers. | 2 | 3 |
| **Hpd index, h per decade** (Kosmulski, 2009) | Compare the scientific output of scientists in different ages. Seniority-independent *h* type index. | 2 | 4 |
| **Hw, citation-weighted h index** (Egghe and Rousseau 2008) | Weighted ranking to the citations, accounting for the overall number of *h*-core citations as well as the distribution of the citations in the *h*-core. | 2 | 4 |
| **hα** (Eck and Waltman, 2008) | Cumulative achievement, advantageous for selective scientists. | 2 | 4 |
| **b-index** (Brown 2009) | The effect of self-citations on the *h index* and identify the number of papers in the publication set that belong to the top n% of papers in a field. | 2 | 4 |
| **hT, tapered h-index** (Anderson et al. 2008) | Production and impact index that takes all citations into account, yet the contribution of the *h*-core is not changed. | 2 | 5 |
| **hrat index, rational h indicators** (Ruane and Tol 2008) | Indicates the distance to a higher *h index* by interpolating between *h* and *h*+1. *h*+1 is the maximum amount of cites that could be needed to increment the *h index* one unit (Alonso et al. 2009). | 2 | 5 |

*h independent indicators, 4b*

Six *h*-independent indicators of cumulative impact were identified, 4 scored a complexity rating of ≤3: The Wu index *w*, *f index*, *g index* and the *t index*. *w* is a simple indicator of prestige, tested in physics and recently economics, that states for example a researcher has a *w index* of 1 if 10 of their publications are cited 10 or more times, but they have not achieved a *w index* of 2 because that implies that 20 of their publications have to have been cited 20 or more times. Wu suggests that *w*1 or 2 is someone who has learned the rudiments of a subject; 3 or 4 is someone who mastered the art of scientific activity, while "outstanding individuals" have a *w index* of 10. The *g index* on the other hand is introduced by Egghe (2006) as an improvement of *h*, as it inherits all the good properties of *h* and takes into account the citation scores of the top articles. *g* claims to provide a better distinction between scientists than *h* as it weights highly cited papers to make subsequent citations to these highly cited papers count in the calculation of the index, whereas with *h* once a paper is in the *h*-core, the number of citations it receives is disregarded. Like *h*, *g* ranks publications after citations in descending order but *g* takes the cumulative sum of the citations and the square root of the sum for each publication. *g* is where the rank and the square root is the same or higher. As such *g* is based on the arithmetic average and ignores the distribution of citations, (Costas and Bordons 2007; Alonso et al. 2009), meaning a researcher can have a large number of unremarkable papers and have a large *g index*. Alternative ways to estimate the central tendency of the skewed distribution of citations to core papers are the *f* and *t* indices. These are based on the calculation of the harmonic and the geometric mean and as such suggested as more appropriate average measures for situations where extreme outliers exist, i.e. the few very highly cited papers. Papers are again ranked in descending order of citations, and beginning with the highest cited paper, the harmonic or geometric mean is calculated stepwise until the product is equal or higher than the rank.

**Table 9.** Indicators that rank publications in the portfolio, *h*-independent indicators.

| ***h*-independent indicators(4b)** | **Designed to indicate** | **Complexity** | |
|---|---|---|---|
| | | **Col** | **Cal** |
| **w index** (Wu 2008) | The integrated impact of a researcher's excellent papers. | 2 | 2 |
| **f index** (Tol 2009) | Attempts to give weight/value to citations. *f* is the highest number of articles that received *f* or more citations on average. | 2 | 3 |
| **g index** (Egghe 2006) | The distinction between and order of scientists (Egghe, 2006; Harzing, 2008). | 2 | 3 |
| **t index** (Tol 2009) | Attempts to give weight/value to citations. *t* is the highest number of articles that received *t* or more citations on average. | 2 | 3 |
| **π index** (Vinkler 2009) | Production and impact of a researcher is computed by comparing the researcher's citation performance "elite" papers ranked top of his or her field. | 2 | 4 |
| **Gα** (Eck & Waltman 2008) | Based on same ideas as the *g index*, but allows for fractional papers and citations to measure performance at a more precise level. | 2 | 4 |
| **Rational g-index (g rat)** (Schreiber 2008a; Tol 2008) | Indicates the distance to a higher *g index*. | 2 | 5 |

*H adjusted to field, 4c*

The indicators in this category claim to adjust different publication and citation habits in different fields and as such present indicators useful for comparing scientists. *Normalized h* is recommended as an adjustment to *h*. It is calculated as the *h index* divided by the number of articles not included in the *h*-core. Meanwhile the *n index* is the researcher's *h index* divided by the highest *h index* of the journals of his/her major field of study. The *n index* is a theoretical index still awaiting validation, and has only been tested using the Scopus definition of *h* and SCImago Journal and Country Rank website for the journal information.

**Table 10.** Indicators that rank publications in the portfolio, *h* dependent indicators adjusted to field.

| **h adjusted to field (4c)** | **Designed to indicate** | **Complexity** | |
|---|---|---|---|
| | | **Col** | **Cal** |
| **n index** (Namazi and Fallahzadeh 2010) | Enables comparison of researchers working in different fields by dividing *h* by the highest *h* of journals in the researcher's major field of study to normalize for unequal citations in different fields. [ | 2 | 2 |
| **Normalized h-index** (Sidiropoulos et al. 2007) | Normalizes *h* to compare researchers' achievements across fields. | 2 | 3 |
| **h index sequences and matrices** (Liang 2006) | Singles out significant variations in the citation patterns of individual researchers across different research domains. | 2 | 4 |
| **hf, generalized h-index** (Radicchi et al. 2008) | Allows comparison to peers by correcting individual articles' citation rates for field variation . | 3 | 4 |
| **x index** (Claro and Costa 2011) | Indication of research level. Describes quantity and quality of the productive core and allows for cross-disciplinary comparison with peers. | 3 | 4 |

*H corrected for co-authorship, 4d*

All the 6 indicators in this category require calculation of the *h*-index in their mathematical foundations. The *alternative h index* or *hi* as it is also known divides *h* by the mean number of authors for each paper in the *h*-core, while *POPh* divides the number of citations by the number of authors for each paper and then calculates *h* using this normalized citation count. Both models give an approximation of the impact authors would have if they had worked alone, however these models treat citations and publications as a single unit that can be evenly distributed.

**Table 11.** Indicators that rank publications in the portfolio, *h* dependent indicators adjusted for co-authorship.

| h adjusted for co-authorship (4d) | Designed to indicate | Complexity | |
|---|---|---|---|
| | | Col | Cal |
| **Alternative h index**, **(also hi)** (Batista et al. 2006) | Indicates the number of papers a researcher would have written along his/her career if they had worked alone. | 2 | 2 |
| **POP variation individual h index** (Harzing 2008) | Accounts for co-authorship effects. | 2 | 3 |
| **Hp, pure h index** (Wan et al. 2007) | Corrects individual *h*-scores for number of co-authors. | 2 | 4 |
| **$h_m$-index** (Schreiber 2008b) | Softens influence of authors in multi-authored papers. | 2 | 4 |
| **$h_{ap}$, adapted pure h index** (Chai et al. 2008) | Finer granularity of individual *h*-scores for number of co-authors by using a new *h*-core. | 2 | 5 |

## Impact over time, category 5

Indicators of impact over time indicate the extent a researcher's work continues to be used or the decline in use. Twelve indicators were identified, six potentially useful, complexity ≤3. Ten indicators were designed to indicate impact over time relative to the portfolio and two allow comparison to the expected aging rate of the field.

*Impact over time normalized to portfolio, 5a*

Eight indicators were identified, ≤3: The age-weighted citation rates (*AWCR*, *AW* and *per-author AWCR*), *AR index, m quotient, mg quotient, Price Index* and *citation age, c(t).* Of these the age-weighted citation rates (*AWCR*, *AW* and *per-author AWCR*), *c(t), m quotient, mg quotient* and *Price Index* are ratio-based models. *AR* is based on the square root of average number of citations per year of articles included in the *h*-core and like the *m quotient* is also *h*-dependent. *m quotient* is the *h index* divided by the length of the researcher's publishing career, which is defined as the number of years since the first publication indexed in the database used to calculate the *h index* to the present year. Similarly, *mg* is the *g index* divided by length of the researcher's publishing career.

**Table 12.** Indicators of impact over time normalized to the researcher's portfolio.

| Impact over time normalized to portfolio (5a) | Designed to indicate | Complexity | |
|---|---|---|---|
| | | Col | Cal |
| **AR index** (Jin et al. 2007) | Accounts for citation intensity and the age of publications in the core. | 2 | 2 |
| **m quotient** (Hirsch 2005) | *h* type index, accounting for length of scientific career. | 2 | 2 |
| **mg quotient** | *g* type index, accounting for length of scientific career. | 2 | 3 |
| **AWCR, age-weighted citation rate, AW & per-author AWCR** (Harzing 2012) | *AWCR* measures the number of citations to an entire body of work, adjusted for the age, *AW* is the square root of *AWCR* to appropriate *h*, and *per-author AWCR* adjusts *AWCR* for the number of authors of each individual paper,. | 2 | 3 |
| **PI, Price Index** (Price 1970) | Percentage references to documents, not older than 5 years, at the time of publication of the citing sources. | 3 | 2 |
| **c(t), citation age** (Egghe et al. 2000) | The age of citations referring to a researcher's work. | 3 | 3 |
| **$h^c$, contemporary h-index** (Sidiropoulos et al. 2007) | Currency of articles in *h*-core. | 2 | 4 |
| **$h^t$, trend h index** (Sidiropoulos et al. 2007) | Age of article and age of citation. | 3 | 4 |
| **Dynamic h-type index** (Rousseau and Ye, 2008) | Accounts for the size and contents of the *h*-core, the number of citations received and the *h*-velocity. | 3 | 4 |
| **DCI**, **discounted cumulated impact** (Ahlgren and Järvelin 2010; Järvelin and Person 2008) | Devalues old citations in a smooth and parameterizable way and weighs the citations by the citation weight of the citing publication to indicate currency of a set of publications. | 3 | 5 |

Inspired by the *AR index* the *AWCR* calculates the sum of citations to an entire body of work, by counting the amount of citations a paper has received and dividing by the age of that paper. The *AW index* is the square root of the *AWCR* to allow comparison with the *h index*, whereas the *per-author age-weighted citation rate* is similar to *AWCR*, but is normalized to the number of authors for each paper. The *Price Index* is the number of citations less than 5 years old from the time the paper was published divided by the total number of citations, multiplied by 100. *c(t)* also indicates the age of citations referring to a researcher's work. A corrective factor is required if citation rates are to be adjusted for changes in the size of the citing population or discipline.

*Impact over time normalized to field, 5b*

Indicators of impact over time adjusted to field are sophisticated indicators, all of which we judged as too complex to be useful to the end-user. *The Classification of Durability* is a percentile based indication of the distribution of citations a document receives each year, adjusted for field and document type. It can detect the possible effects durability can have on the performance measurement of an individual. However, at the present time its analysis is limited to WoS. *a(t)* corrects the observed citation distribution for growth, once the growth distribution is known. Costas et al. (2010a) propose combining indicators to produce classificatory performance benchmarks. Their indicator *Age and Productivity* combines the mean number of documents by age and cites per paper (using a three year citation window) in four year age brackets adjusted to field to identify the age at which scientist produce their best research and to some extent the decline in their knowledge production. But the demanding data collection, multiple indicators and dependence on WoS journal categories make it unlikely that an end-user will take the time needed to calculate the indicator.

**Table 13.** Indicators of impact over time normalized to field.

| **Impact over time normalized to field (5b)** | **Designed to indicate** | **Complexity** | |
|---|---|---|---|
| | | **Col** | **Cal** |
| **Classification of Durability** (Costas et al. 2010a; 2010b; 2011) | Durability of scientific literature on the distribution of citations over time between different fields. | 2 | 4 |
| **a(t), aging rate** (Egghe et al. 2000) | Aging rate of a publication. | 3 | 4 |
| **Age and Productivity** (Costas et al. 2010a) | Effects of academic age on productivity and impact. | 3 | 4 |

## Discussion

Our initial analysis of the indicators highlighted two problems: 1) The availability and accessibility of publication and citation data does not support the practical application by end-users of the indicators. Many indicators are invented for ideal situations where complete datasets are available and do not cater to real life applications. 2) Some indicators lack appropriate validation and recognition by both the bibliometric and academic community. Judging by the quantity and availability of the indicators we identified, it is obvious that end-user bibliometric assessment has the potential to go beyond the *h index* and *JIF*. This paper has only focused on the effects of publications, which is a small area of scientific activity. Still, even for this one activity, sub-dividing indicators of "effects of publications" into different aspects illustrates how essential it is to recommend groups of indicators to end-users rather than single indicators. Presenting indicators in categories is a way to demonstrate how different aspects of performance can be captured, as each indicator has its own strengths and weaknesses as well as "researcher/field" variables that can be redundant or counter-productive when indicators are used together. Even though our schematic presentation simplifies understanding what the indicators do, recommending useful indicators is still a challenge. The benefit of choosing one over the other is highly dependent on the spread of the end-user's publication and citation data, the academic age of the end-user and the availability of the data. In the following we discuss some of the main issues for each category.

### Publication Count, category 1

Indicators of publication count provide information of the sum of a researcher's publications produced within a given timeframe, Table 3. We judged all 15 indicators useful for end-user application however they are some limitations that users of these count-based indicators need to be aware of. Count alone provides a distorted picture of the scope of a researcher's output and divulges nothing about the level of contribution to a work unless authorship credit is explicitly stated (Hagen 2010). In the assessment of contribution, validation is required from all authors of actual contribution to a paper, as name order in the by-line can be strategically or politically motivated or simply alphabetical (Bennett and Taylor 2003). If it is normal for the discipline to have many authors per paper rather than single authored papers, correcting for single author contribution is superfluous and perhaps counterproductive. Count can be balanced by weighting different forms of publication, be it patents, books, book chapters, articles, enlightenment literature, conference papers etc., after importance for the field in which the researcher is active. Though which document types and how they are weighted needs to be clear. The value given to a specific type of publication varies from discipline to discipline but on an individual level could be weighted in relation to the mission and resources of the researcher's affiliated institute. Weighting output types should, however, be used with caution as the positive or negative effect this has on publishing behaviour needs further investigation. Also, weighting can make the comparison to normalised national and international standards unreliable as document type has to be compared with the exact same document type, which can result in the preference of some forms of publications to the detriment of others in the computation of the standards.

### Qualifying output as Journal Impact, category 2

Journal and article impact indicators are frequently added by end-users to CVs next to publications as a proxy for the level of quality of a published paper. In assessments for jobs or tenure they are used as a selection parameter to judge applicants' publications and as benchmarks for expected disciplinary performance. They were originally designed to indicate the average impact of articles published in journals in a defined publishing year and with a short citation window or to aggregate the publications of a research group or center. The journal-based citation indicators in Table 4 are dependent on journal performance and have been shown to measure popularity and not prestige. Popularity is not considered a core notion of impact (Bollen et al. 2006; Bollen and Sompel 2008; Yan and Ding 2011). As such they are dependent on the disciplinary characterisation of publications and citations, journal aggregation in sub-disciplines in citation databases, the methodology used to estimate citations and the type of papers included (excluded) in the calculation. However, where the individual publishes is considered an important criterion in the assessment of visibility and impact. Yet, as Table 4 illustrates, the construction of the impact factors means they are an indication of researchers publishing success and not the actual use of their articles. In the investigation of the use of journal impact factors at the author-level it is necessary to study if time and impact of journals correlate in the same way in the assessment of individual impact. Our results identified only one indicator of impact designed for assessment at the author-level and simple enough for the researcher to use; $P_{tj}$ (articles published in journals deemed relevant or prestigious by heads of department or institution). *Ptj*, can of course be extended to encompass other types of publications, as to support non-journal based fields, and can also be extended to source deemed authoritative by other that heads of department.

## Effect of Output, category 3

Indicators of the effect of output can be grouped into three types of aggregation: number of citations, averages or percentiles. Calculations in all approaches are relatively simple but in practice the availability of data makes the feasibility of the end-user using these indicators to produce reliable indications of the true effect of the publications questionable. As field coverage is limited in citation databases, citation indicators are more appropriate in some fields than others. Ideally citation indicators require data collection in multiple sources to provide as complete a picture of "use" as possible, however, the overlap between citation databases requires the end-user to filter out duplicate citations. This immediately adds to the complexity of the indicators.

*Effect as citations, 3a*

Citations are counted as the amount of times a paper has been used in other published academic papers. For our recommended 9 indicators, the count is limited to citation databases that index citations, e.g. WoS, Scopus, Google Scholar and the resulting count can differ from database to database. Further the count can differ between versions of the same database, dependent on the subscription of the end-user. Count does not reveal if the citations have been positive or negative, the currency of the citations or if the count is due to older articles having more time to accumulate citations. Citations can be interpreted however as the contribution of research to the social, economic and cultural capital of academic society and /or an indication the interaction between stakeholders, how new approaches to science are stimulated, and influence on informing academic debate and policy making (Directorate General for Research 2008; Bornmann 2012). Consequently, a high citation count is desired and the indicator *MaxC* is a proxy for the researcher's most prestigious paper. Likewise the *i10 index* indicates substantial papers. Putting the arbitrariness of a minimum of 10 citations as a cut-off point to one side, the index is based on a Google Scholar, whose database content is not transparent and suggested vulnerable to content spam and citation spam (Jascó 2011; Delgado Lopéz-Cózar et al. 2014). To understand if the citations are to a few papers out of the end-users entire portfolio, the *nnc* indicator counts the number of papers that have not received any citations. *nnc* is not an indicator of low quality work, but is a useful indicator that helps interpretation of other performance indicators that build on average citations per paper. Publications can be greatly used and of great influence, but never cited. Certain types of publications are important but rarely cited such as technical reports or practice guidelines. Lack of citations can be caused by restricted access to sources, fashionableness of the topic, changes in size of citing or citable population and the citability of different types of publications (Egghe et al. 2000; Archambault and Larivière, 2010; Costas et al. 2010b).

Scientists build on previous findings, so self-citation, *sc*, is unavoidable. Excessive self-citations inflate citation count and are considered vanity and self-advertising. In assessment self-citations can affect the reliability and validity of citation count on small amounts of data (Glänzel et al. 2006; Costas and Bordons 2007). However, there is no consensus in the bibliometric community if removing self-citations has any effect on robust indicators or if the removal process can introduce more noise in the citation count than is removed (Harzing 2012). Most citation indexes have the option to remove self-citations but what constitutes a self-citation is undefined, as they can be understood as citations by the researcher to own work, citations from co-authors of the paper or citations from a colleague in the same research group. Alternative indications of the importance of scholarship share the citations between researchers that have contributed to a paper. *Fractional citation count*, i.e. averages - geometric, harmonic and arithmetic - are affected by the skewed distribution of citation data which is why there is a movement in the literature towards the stability and consistency of percentiles (Belter 2012). Consequently, is has been recommended not just to compare results obtained from several databases, but combine citation counts with other methods of performance assessment and only then normalise results of individual performance to academic seniority, active years and field to ascertain excellence (Costas et al. 2010a).

*Effect as citations normalized to publications and field, 3b*
Percentiles such as *E(Ptop), A/E(Ptop), Ptop* are considered as the most suitable method of judging citation counts normalized in terms of subject, document type and publication year as they attempt to stabilise factors that influence citation rates (Bornmann and Werner 2012). Bornmann argues for their simplicity of calculation, which is debateable, but they are more intuitive to the end-user than average cites-per-paper in that visualization of results in box-charts or bar-charts can provide easy-to-read presentations of performance. Percentages have the further advantage that they are only affected by the skewed distribution of citation data to a limited extent and are adjustable to individual assessments as measures of excellence. *Ptop*, for example, can be adjusted to *Ptop/author* to illustrate the amount of papers a scientist has within the top 5% papers within a field, and as such indicate excellence (van Leeuwen et al. 2003). Field indicators, *Field Top %,* favour some fields more than others; older articles, senior scientists with extensive publishing careers and are often based in predefined journal-subject categories in citation databases. The degree to which top n% publications are over or under-represented differs across fields and over time (Waltman & Schreiber 2013). Likewise the indicator *Significant Papers* adjusts the number of citations that are considered significant for seniority, field norm and publication types, which results in a subjective indicator that can randomly favour or disfavour researchers. Data-completeness, differences in citation rates between research fields, and the need for a sufficiently large publication output to obtain a useful percentage benchmark at the author-level compromise the simplicity and stability of these comparative measures of excellence. Hence they may not be representative of the response to a researcher's work, but they can prevent a single, highly cited publication receiving an excessively heavy weighting.

To interpret individual researcher impact, end-users compare themselves with peers to understand the level of their performance, however using field normalization to cater for different publication and citation traditions is not without its difficulties. It means that up-to-date reference standards for the field have to be available to the end-user. Reference standards fix the field by calculating normalizing factors using multiplicative correction and other parameters (Iglesias and Pecharromán 2007). Studies have shown that normalized indicators characterise the area but can be disadvantageous for the specific publication patterns of researchers within their sub-field specialty (Ingwersen et al. 2001; van Leeuwen and Moed 2002; Bollen et al. 2006; Yan and Ding 2011). Further, normalization favours highly cited researchers as impact increases in a power law relationship to the number of published papers (Iglesias and Pecharromán 2007) and assume that publication and citations are independent variables. In other words the effect of the publishing size on the citation count has been eliminated. Using journal subject categories is an accessible way to define fields, but it is doubtful if researchers can feasibly indicate their global impact using journal impact-defined field indicators as these are normalized to a field that neither accurately represents the specialty of the researcher or individual researcher demographics, such as seniority or academic age. Antonakis et al. (2008) propose the *IQP* indicator to enable researchers to compare the performance of their papers to other papers within their specialty. *IQP* produces descriptive indicators of the global number of citations a researcher's work would receive if it were of average quality in its specialty, by calculating the ratio actual citations to estimated citations using the journals the researcher publishes in as a proxy for "specialty". As a result the end-user can indicate the number of papers (corrected for subject and academic age) which perform above the expected average for the specialty and how much better than average these papers perform. Acknowledging how time consuming the indicator can be to calculate for the end-user, they provide a free online calculator and benchmarks for interpretation[3]. The indicator has in tests correlated better with expert ratings of excellence than the *h index*. The indicator is again dependent on subject categories and citation record in WoS which makes the *IQP* more useful only to researchers well represented in WoS.

High scientific quality is not necessarily related to high citation count, but perhaps most important for assessment is to acknowledge that the true impact of a piece of research can take many years to become apparent and the routes through which research can effect behaviour or inform social policy are diffuse. Therefore we include in our analysis Neiderkrotenhaler et al.'s questionnaire tool (2011). The tool indicates the broader impact of publications by combining the interest of societal stakeholders with quantitative indicators of knowledge dissemination and use. It assesses the effect of the publication in non-scientific areas, the motivation behind the publication and efforts by researchers to disseminate their findings.

*Effect as citations normalized to publications in portfolio, 3c*

The average (mean) cites per paper *CPP*, or *medianCPP* are robust measures for comparisons of researchers to field averages or comparisons between researchers who have been active for different numbers of years. The mean and median are different measures of the central tendency in a set of data, or the tendency of the numbers to cluster around a particular value. In bibliometrics it is desirable to find the value that is most typical. One way of doing this is to find the mean, or average, which is the sum of all the citations divided by the total number of publications. Another way is to find the median, or middle, value, which is the one in the centre of an ordered list of publications ranked after the amount of citations they have received. The average has the disadvantage of being affected by any single citation being too high or too low compared to the rest of the sample. *CPP* seems to reward low productivity. This is why *medianCPP* is taken as a better measure of a mid-point or percentiles are preferred. Percent self-citations *%SELFCIT*, percent non-cited publications after a certain time *%PNC* and percent not cited over all publications *%nnc* are ratio-based indicators which account for the lack of citations or lack of external citations relative to the amount of publications in the portfolio.

The currency of publications can be analyzed by looking using *Age of citations*. This indicator predicts the useful life of documents over a period of time. Moreover, it helps end-users select the significant (most used) papers and understand how their papers are used – if older papers have first come of age recently and are accumulating citations, if their papers have a short "shelf life" or if they are constantly used.

**Indicators that rank the publications in the researcher's portfolio, category 4**

The indicators in the following categories are an expression of cumulative impact in a single index, as they calculate the quantity and impact of articles into an indication of prestige (Hirsch 2005; Schreiber et al. 2012). To do this comprehensively, the majority are recommended, by their creators, to be combined with other indicators. When used alone the indicators give only a rough measure of quality as the correlation between output, quality and impact remains uncertain (Nederhof and Meijer 1995; Haslam and Laham 2009). To overcome these shortcomings, "quality" is assumed a value of citation count, as a large number of citations are interpreted as "usefulness" to a large number of people or in a large number of experiments. Our results show that attempts to improve *h* can be at the cost of simplicity and usability, Tables 8, 10, 11 & 12. The descendants of *h* are supposedly more precise, yet in many cases their consistency and validity remains problematic. Some have performed well in laboratory studies: *b* (Brown 2009), *IQP* (Antonakis and Lalive 2008), *h index sequences and matrices*, (Liang 2006), while others have faltered: *h*, *g*, *r*, $h^2$ (Waltman and van Eck 2009). Of course the indicators that incorporate *h* in their foundations suffer from the same inconsistencies as *h*: *hg*, $Q^2$, *normalized h*, *hrat*, *grat*, *a*, *hw*, *ħ*, *e*, *hpd* and *hmx*. Others give undue weight to highly cited papers, *h*, *f*, *t*, *w*, $h^2$ (Schreiber, 2010) and although some of the sampled the indicators proclaim higher accuracy and granularity, these benefits are lost on the end-user as the complexity of the calcuations mean usability and transparency are reduced. *hα*, *Hpd*, *hw*, *hrat* require multiple and advanced calculations, while *hT* requires special software for computation.

[3] The IQP calculator can be downloaded from: http://tinyurl.com/nj7s834

*h-dependent indicators, 4a*

The *h index* already plays an important role for end-users (Costas and Bordons 2007) and despite its flaws, is unavoidable as its simplicity and recognisability outweigh debates of its representativeness. Generally, *h*-type indicators are estimated as stable once a scientist has reached a certain level of scientific maturity, >50 papers, otherwise stability issues can lead to misleading results. The exponential growth of the number of papers advocating the advantages and hazards of the *h index* makes it impossible to present a complete reference list. Briefly, the *h index* has been criticised for negatively influencing publication behaviour (Egghe 2006; Harzing 2008), reducing validity in cross-domain comparison and bias towards certain fields (Podlubny 2005; Iglesias and Pecharromán 2007), having granularity issues, (Vanclay 2007; Harzing 2008), losing citation information (Waltman and van Eck 2011), under-estimating the achievement of scientists with selective publication strategies, women and researchers who have had taken a break from academia, as well as favouring seniority (Costas and Bordons 2007). Perhaps, most importantly, is the questionable arbitrariness of the *h* parameter (Alonso et al. 2009). Subsequently, the indicators that build on the *h index* suffer the same inadequacies as *h*. All of these criticisms must be known outside of the bibliometric community to produce *informed* end-user assessment.

To compensate for limitations of single indicators, we recommend combining *h*-type indicators, however information redundancy is an issue, as investigated in Panaretos and Malesios (2009) and Bornmann et al. (2011). Their investigations reveal high inter-correlations between the *h*-type indicators and they conclude that the various indicators can be redundant in empirical application. Separating the indicators into categories “fundamental” and “derived” reduces the chance of information redundancy in assessments (Zhang 2009) where, for example, *A* and *R*, are *h*-dependent (derived) and thus have information redundancy with *h*. Both Bornmann et al. (2008) and Schreiber et al. (2012) recommend a more user-friendly approach, which is to categorize and combine pairs of indicators relating to the productive core. Using our identified indicators we recommend combining one of the following indicators of the productive core: *h*, *m*, $Q^2$, $h^2$ or *ħ*, with an indicators relating to the impact of papers *A*, *R*, *AR*, *m* or *e* to produce insightful results.

*h independent indicators, 4b*

The *g index* is based on the arithmetic average which means it ignores the distribution of citations, Table 9 (Costas and Bordons 2007; Alonso et al. 2009). However the arithmetic mean has the disadvantage that it is disproportionate to the average publication rate meaning that the *g*-index of a scientist with one big hit paper and a mediocre core of papers could grow in a lot comparison with scientists with a higher average of citations (Tol 2009). Attempts to improve the desirable mathematic properties of the *g index* are the *f* and *t* indicators that use the harmonic and geometric mean. These claim to improve discrimination between similar scientists as *f* weights the distribution of citations and *t* is even less effected by highly-cited papers than *f*. Yet in the broad ranking of researchers calculating the *g*, *h*, *f* and *t* indicators, adds more work but no greater insight in a researcher’s performance (Tol 2009). *h* is always $\leq f \leq t \leq g$, similarly, Glänzel and Schubert (2010) suggest using the median of the citations within the core, the *m index*, and show the *m index* and the *f index* to be less affected by outliers than the other measures. *m* is simpler to calculate than *f* and *t* and results in the same or very similar index number. The *Gα* and the *rational g index* allow for fractional papers and citations to measure performance on a more precise level, however they require setting a value of α and interpolating between *g* and *g*+1 based on the piecewise linearly interpolated citation curve. Consequently, we scored them too complicated for the end-user to use.

Completely independent of the construction issues of *h* and *g* is the *w index* (Wu 2008). The *w index* is a useful and simple way to assess the integrated impact of a researcher's work, especially the most excellent papers[4].

*h adjusted to field, 4c*

Field variation creates obstacles to fair assessment of scientific performance. The *n index* and *normalized h index* have been specifically designed for across field comparison and account for the multidisciplinary of researchers, Table 10. Even though these are simple to calculate, they have some severe limitations. The *n index* divides *h* by the highest *h index* of the major journal the researcher publishes in. In many cases, the *h index* will be based on articles in different areas of science and can have no relation to the highest *h index* of the journals of his major field of study, making the calculation impossible. The *normalized h* can only be used in parallel to the *h index* and rewards less productive but highly cited researchers. Other alternatives are the *x, hf, h index sequences and matrices* indicators but these require advanced multiple calculations, special software and the determination of cut-off values, parameters, stretching the exponential distribution to fit the dataset or field characteristics. These approaches increase confusion over which data is included in the calculation and how it is calculated. If information is lost during the data manipulation the validity is challenged.

A simple option suggested by (Arencibia-Jorge et al. 2008) is to combine *h* type indicators, *h, g* and *A,* to establish quality benchmarks at a lower level of aggregation than the field. They suggest computing successive *h* type indices to account for performance on a "researcher:department:institution" hierarchy.The ranking of researchers at these three levels allows the evaluation at the micro-level, identifies researchers with higher than expected impact as well as aggregated departmental and staff behaviour within the institution and international visibility. Although their solution is interesting, the complexity of data collection increases with the hierarchy and as the indicator becomes a tool for institutional evaluation rather than author-level performance we have not included it in our list of recommended indicators.

*h corrected for co-authorship, 4d*

Assessment of co-authorship is important for the individual researcher in assessment because research collaboration lies at the heart of expressing research activity, knowledge advancement and communication. Simple indicators of *h* adjusted for co-authorship shouldn't be difficult to calculate because the researcher should have all the necessary information - who wrote the articles and their affiliation during publication; homonyms of author and institute names; and the relation between authorship order and contribution. Normalising the *h index* for multi-authorship, (*hi, POP variation, n, hm, alternative h, pure h,* and *adapted pure h*), immediately affects the simplicity of the calculation of *h*. *h* indicators adjusted for co-authorship are calculated in two ways: if the citation count is normalized to the amount of co-authors before or after the h-calculation. For instance, increasing the numbers of papers in the *h*-core affects the precision of the indicator, as in *hm*, while reducing the amount of papers in the *h*-core, *hi*, makes the results sensitive to extreme values and discourages collaborations that can result in multi-authored, highly cited and influential papers.

[4] The calculation is explained above table 9.

It is unclear which indicator is best. Egghe et al. (2000) argue that one particular method of adjusting for co-authorship does not contain an absolute truth and that therefore it is unclear which distribution of the credit to co-authors is the correct distribution. In reality authorship can be rewarded as part of departmental publishing deals, or even as a thank you for permission to access data. We will not be discussing "political" authorship agreements in this review, but from the end-users' point of view the desirability of correcting for co-authorship is doubtful as recalculation of the *h*-core can lead to over-correction and thus penalise the researcher under assessment (Rosenberg 2011). The recurring question is, if sharing credit is at all necessary. Realistically, we expect end-users to present the highest number of citations their works have achieved or the highest scoring indicator. If all researchers within a field practice "multiple co-authorship" then sharing the credit is superfluous and in some cases counterproductive. Not only will researchers reduce their performance on their CV, their *h*-indicators will also be reduced. More importantly, future participation in collaborative projects could be discouraged. So even if we agree that harmonic counting gives a more accurate assessment of collaborative scientific productivity and counterbalances the biases of equalization and inflation when issuing author credit (Hagen 2010), it is worth considering if, within the practices of the field, the extra effort is at all necessary.

## Impact over time, category 5

*Impact over time normalized to portfolio, 5a*

It is incorrectly assumed that the chance of a researcher's work being used declines with age because in general its validity and utility decline as well. Usage and validity are not related, and linking usage with validity is unwise (De Bellis 2009). The rate of loss of validity or utility of older documents is not the same in all fields and does not have to same effect on usage. Literature in the natural sciences ages more quickly than literature in the humanities where information in older documents is more readily incorporated elsewhere. Stochastic models allow for the translation of diverse factors influencing aging into parameters that can be estimated from empirical data with a specified margin of error; *Dynamic h*, *AWCR*, *AW*, *DCI*, $h^t$ (De Bellis 2009). However the calculation of ratio or percentile based models are simpler to understand; *c(t)*, *aging rate*, $h^c$, *m quotient*, *Price Index*, *AR*. Obviously, in these simpler models, the yard stick measure of expected performance is rougher and the illustrated decay of a publication is in some cases steeper, e.g. *AR index*.

*Impact over time normalized to field, 5b*

The more a field grows the more articles come into existence, acting as competition between "older" articles to get into the reference list of the new ones. Growth has been verified as an influence on impact over time but is not a cause of the obsolescence of publications (Egghe and Rousseau 2000). Therefore, if publications from particular researchers need more time than "normal" to be properly acknowledged by their colleagues, the impact of these researchers may be underestimated with standard citation windows. The rate at which scientific literature ages and the rapidity with which it is cited are important in determining the length of the citation windows used for citation counts making field comparisons complex. Measures of impact over time have to cope with diverse characteristics and fluctuations in usage by local groups. The relative or expected (probabilistic) number of citations an individual article receives over an analyzed time interval, adjusted to the local field and document types, are relevant indicators of sustainability at the author-level. Even though the resulting indicators are more nuanced and allow for a greater granular comparison of research performance over time, we judged the measures too complex for end-user application.

### Methodological considerations

This review is limited to a subjective complexity assessment of indicators at the individual level. Our judgements perhaps underestimate the abilities of end-users, especially the end-users that practice using bibliometric indicators and are very knowledgeable about their limitations. Our search for indicators taught us that some researchers are very keen on using bibliometric indicators on their CVs and include a narrative explaining the computations of the indicators they use. Some have gone so far as inventing their own domain specific indices. We have not tested empirically the complexity of each indicator neither have we investigated the applicability, validity, utility, objectivity or the effects on publishing behaviour. Further we have not studied the cause and effect mechanisms inherent to the indicator, or inter-field variations of the indicators when implemented. Neither have we considered the reliability of indicators used by end-users on their CVs. These need to be analysed in future studies involving end-users.

The categorization of indicators covers the basic effects of publications at the author-level. Our simple set of categories, even if they do not converge with other typologies, provides valuable information on the relative merits and weaknesses of the indicators. The qualitative approach was preferred as comprehensive factor analysis was not the purpose of this review.

## Conclusions

We did not identify a single indicator that captures the overall impact of a researcher. Our categorization illustrates clearly that author-level indicators only partially capture individual impact as they indicate impact over time, impact normalized to field, impact of a selected number of publications or impact normalized to the researchers' age, seniority and productivity. Only when indicators are used in combination can they approximate the overall impact of a researcher. Hopefully our review will increase awareness of the range of options end-users have to demonstrate the impact of their work and will discourage using a single numerical value to represent the effects of their work. However, choosing the appropriate indicators to combine takes knowledge of which aspect of publication activity the indicator attempts to capture and how the indicators are calculated, including which data needs to be collected. As there is no workable definition of scientific impact, there is no agreement on which combination of indicators best express the impact of a researcher's body of work or which best fit the aim of an assessment of a researcher. But there is at least agreement that using just one indicator is inadequate.

Administrators, evaluators and researchers seem to use indicators as never before and their widespread use has led to the construction of novel indicators as well as variants or combinations of established ones. This paper reviews 108 author-level indicators and exemplifies the complexity of their data collection and calculation in relation to end-user application. Our study attempts to identify which author-level indicators can be calculated by end-users and we succeeded in identifying 79 such potentially useful indicators. The data collection and calculation of these indicators is relatively straightforward, and as such it is clear how they measure or interpret certain aspects of performance. Further, our study shows that superior author-level indicators that claim to produce improved representations of individual performance and more granular distinctions between researchers, were too complicated for end-users to apply them in practise.

As indicators get more refined their complexity increases and as such we assume they are designed for the bibliometric community to use in assessments on the behalf of the individual and not for end-user "self-assessment". The results show that at the current time 1) certain publication activities and effects are more easily evaluated using bibliometrics than others, 2) assessment of publication performance cannot be represented by a single indicator, and 3) it is unwise to use citations as anything other than an *indication* of impact. Our clarification of how the indicators are calculated clearly demonstrates that the majority of indicators are different approximations of the average citations to publications in a dataset. Which indicator is the best approximation of the average is dependent on the data used in the calculation. To choose the best indicator, the end-user has to understand the spread of the data and which indicators present the best model that captures the central tendency. However, unlike statistical models the indicators produce solitary numbers as an estimate of performance, which are presented to the end-user without confidence intervals or minimum/maximum values that would provide contextual information about these point estimates.

Bibliometric indicators are readily available, and will therefore be used in both intended and unintended ways. Using indicators out of their context is a problem in relation to their validity or rather the validity of the use made of the measure. Which indicators are most useful to an end-user in expressing their publication performance requires further study. Taking one indicator alone and interpreting the results out of context of the researcher's field or seniority will result in distorted and useless information. We can conclude that by providing a recommended selection of indicators for end-user assessment, the researcher can reach a better understanding of the impact of their published works and perhaps identify where this can be improved. The success of the indicators are dependent on the completeness of data, which often requires access to comprehensive citation databases and the extraction of unstructured data from the internet or other sources. The knowledge we have about which indicators individuals can employ to reliably measure their performance is limited. They have yet to be properly validated using empirical data from different research fields and their long term effects on scientific behaviour needs to be investigated in prospective studies. However, our extensive tables can contribute to awareness of the possibilities and limitations of bibliometric indicators as well as the data requirements, time and competencies needed to calculate them. Simple indicators are recommended for end-user application as their requirements to bibliographic data are modest and calculations transparent.

## Acknowledgements

This work was supported by funding from ACUMEN (Academic Careers Understood through Measurement and Norms), FP7 European Commission 7th Framework "Capacities, Science in Society", grant Agreement: 266632. Opinions and suggestions contained in this article are solely the authors and do not necessarily reflect those of the ACUMEN collaboration.

# Appendix

## Publication Count, category 1: *All indices require verified publication data.*

| Indicator | Definition | Designed to indicate | Individual | | Complexity | | Comments |
|---|---|---|---|---|---|---|---|
| | | | Advantages | Limitations | Col.* | Cal.* | |
| **P** | Total counting. Each N author of a paper receives 1 credit. | Count of production used in formal communication | Potentially, all types of output can be included or selected in regards to theme of evaluation. | Does not measure importance, impact of papers, duration or volume of research work. | 1 | 1 | Counts vary across disciplines due to nature of work and conventions for research communication. |
| **FA** First author counting | Only first of N authors of a paper receive a credit equal to 1. | Credit given to first author only | Simple method of crediting publication to the assumed main contributor. | Does not give an accurate picture of the relative contribution of the authors | 1 | 1 | Unfair when authors are ordered alphabetically or practice 'noblesse oblige' (Russell and Rousseau 2002) |
| **Weighted publication count** | Applies a weighted score to the type of output. | A reliable distinction between different document types. | Accounts for importance of different publication types for communication within a field. | Has to be designed individual to field as no gold standard. | 1 | 1 | Enables comparisons of like with like. |
| **Patent applications** (Okubu 1997) | Count of patent applications | Innovation | Resources invested in R&D activities and role of scientist in development of new techniques. | Patent application varies from field to field. | 1 | 1 | Quality or significance of patents is not on an equal level; |
| **Dissemination in public sphere** (Mostert et al. 2010) | Count of contributions to, inc.: tv & radio pro-grams, newspapers, non-peer reviewed journals, text books, public & professional websites and news forums | Impact and use in public sphere (knowledge transfer) | Useful addition to evaluation of scientific dissemination activities in the academic environment; | Many indicators and no gold standard method of weighting relative to departmental norm or expected performance in discipline | 1 | 1 | Societal quality is dependent on different activities than scientific quality and is not a consequence of scientific quality. |
| **Co-publications** | Count or share of co-authored publications. | Collaboration on departmental, institutional, inter- or national level & identify networks. | Shows with whom researcher co-publishes and the intensity of co-publication | Usefulness is affected by how the identification of affiliation and partnerships is handled. | 1 | 1 | Identifies if collaboration is governed by immediate proximity. |
| **Number of co-authors** | Count of authors per paper | Indicates cooperation and growth of cooperation at inter- and national level; | Measure volume of work by teams of authors at individual level | Whole or fractional counts of authorship produce different results | 1 | 1 | How affiliation is listed can be problematic and can affect aggregation. |
| **P (publications in selected databases) fx** $\mathbf{P_{isi}}$ | Number of papers in ISI processed publications | Used in the calculation of impact compared to world subfield citation average based on ISI citation data. | Recognised benchmark for analyses and bibliometric research projects. | Includes only ISI defined normal articles, letters, notes, reviews and conference papers. | 1 | 2 | Provides a distorted or incomplete picture; more appropriate in some fields than others (Harzing 2012). |
| $\mathbf{P_{ts}}$ | Publication in selected sources | Number of publications in selected sources defined important by the researcher's affiliated institution. | Reflects output in sources deemed locally important. | Provides only a snapshot of productivity | 1 | 2 | Provides a distorted or incomplete picture |

# Appendix

| | | | | | | | |
|---|---|---|---|---|---|---|---|
| **Fractional counting on papers** | Each of the N authors receives a score equal to 1/N | Shared authorship of papers gives less weight to collaborative works than non-collaborative ones. | Accounts for differences in publishing behaviour among fields of science and level of multi-authorship. | Favours secondary authors by allocating equal credit to all authors | 1 | 2 | Criticized for lack of fit between credit scores and contribution (Hagen 2010) |
| **Proportional or arithmetic counting** | Author with rank R in by-line with N co-authors (R=1,..N) receives score N+1-R | Shared authorship of papers, weighting contribution of first author highest and last lowest. | Rewards level of contribution to a paper. | If authors adapt alphabetical ordering or take turns to be first or second author this indicator cannot be applied. | 1 | 2 | Can be normalized in such a way that the total score of all authors is equal to 1. |
| **Geometric counting** | Author with rank R with N co-authors re-ceives credit of 2N-R | Assumes that the rank of authors in the byline accurately reflects their contribution | The first few authors get most of the credit | Allotted authorship credit rapidly approximates asymptotic values as N increases. | 1 | 2 | Asymptopic values lose their validity on small sample size. |
| **Harmonic counting** | Ratio of credit allotted to ith and jth author is j:i regardless of total number of co-authors | The 1st author gates twice as much credit as the 2nd, who gets 1.5 more credit than the 3rd, who gets 1.33 more than the 4th etc., | Provides accurate representation of perceived quantitative norms of byline hierarchy. | Applies only in areas where unequal co-author contributions are the norm. | 1 | 2 | Tested in natural sciences |
| **Noblesse oblige** | Last author gets 0.5 credit, other N-1 authors receive 1/(2(n-1)) each | Indicates the importance of the last author for the project behind the paper. | Acknowledges that the last author contributes with resources and not data | There is no way to identify actual level of contribution apart from statements from the authors. (Bennett & Taylor 2003) | 1 | 2 | This is one of many suggested counting schemes for noblesse oblige |
| **Cognitive orientation** | Analysis by aggregating papers according to scientific subfields the individual publishes or is cited in. | Identify how frequently a scientist publishes or is cited in various fields; indicates visibility/usage in the main subfields and peripheral subfields. | Can easily be related to the position a researcher holds in the community | More applicable in some fields than others as often journal based and limited to CI† definition of scientific fields | 2 | 1 | Useful to identify future areas for collaboration and production. |

* Col. = data collection, Cal. = calculation

# Appendix

## Qualifying output as Journal Impact, category 2

*All indices require verified publication data and data from one or more citation databases. Some require an aggregate of "world" publication and citation data to calculate field normalisation scores.*

| Indicator | Definition | Designed to indicate | Individual | | Complexity | | Comments |
|---|---|---|---|---|---|---|---|
| | | | Advantages | Limitations | Col* | Cal* | |
| **ISI JIF (SIF) Synchronous IF** | Number of citations a publication has received during a single citing year to documents from previous 2 publication years | Average number of citations a publication in a specific journal has received limited to ISI document types and subject fields. | Readily available. The "mix" of different publication years makes SIF robust indicator of permanent impact | Measure of journal popularity not scientific impact (Bollen et al. 2006). Not designed for indication of individual performance. | 2 | 1 | Does not allow for different citation window to benefit field; hides variation in article citation rates as citations are results of skewed distribution. |
| **SNIP (source normalized impact per publication)** (Moed 2010;Waltman 2013) | The number of citations given in the present year to publications in the past three years divided by the total number of publications in the past three years | The citation impact of scientific journals using a so-called source normalized approach. Normalization of citation counts for field differences are based on characteristics of the sources from which citations originate | Citations are normalized in order to correct for differences in citation practices between scientific fields. | source normalization does not correct for differences between fields in the growth rate of the literature or unidirectional citation flows from one field to another (e.g., from an applied field to a more basic field) | 2 | 1 | Revised by (Waltman 2013) to correct counterintuitive properties. Does not require a field classification system in which the boundaries of fields are explicitly defined |
| **Immediacy index** | Ratio number of citations a journal receives in a given year to the number of articles it issues during the same year. | Speed at which an average article in a journal is cited in the year it is published | Discounts the advantage of large journals over small ones. | Frequently issued journals may have an advantage because an article published early in the year has a better chance of being cited than one published later in the year. | 2 | 1 | Different types of journals influence the immediacy index, such as length of publishing history, prestige and atypical references. |
| **Aggregate Immediacy Index (AII)** | AII cites to all items published in journals in a particular subject category in one year divided by the number or articles/reviews published in those same journals in the same year | How quickly articles in a subject are cited | Useful context for evaluating how a journal compares to other journals publishing within the same discipline. | Metric can be limited by field coverage of citation database. | 2 | 1 | For comparing journals specializing in cutting-edge research, the immediacy index can provide a useful perspective. |
| **Cited half-life (CHL) & Aggregate Cited Half-Life (ACHL)** | CHL is the number of years, going back from the current year, that account for 50% of the total citations received by the cited journal in the current year | A benchmark of the age of cited articles in a single journal | ACHL is an indication of the turnover rate of the body of work on a subject and is calculated the same way as CHL. | A lower or higher cited half-life does not imply any particular value for a journal. | 2 | 1 | It is possible to measure the impact factor of the journals in which a particular person has published articles however misuse in evaluating individuals can occur as there is a wide variation from article to article within a single journal |

# Appendix

| | | | | | | | |
|---|---|---|---|---|---|---|---|
| **IFmed** (Costas et al. 2010a) | IF med is the median value of all journal Impact Factors in the subject category. | The aggregate Impact Factor for a subject category | Accounts for the number of citations to all journals in the category and the number of articles from all journals in the category. | The number of journals that make up categories and the number of articles in these journals influence the calculations of these ratios. | 2 | 2 | Not designed to replace the JIF, but is a complementary indicator. |
| **Ptj** (Rehn et al. 2007) | Count of number of publications published in selected journals in a time span. | Performance of articles in journals important to (sub)field or institution. | Reflects potential impact of articles in sources defined locally as important. | Does not take the size of the analyzed unit into account. | 1 | 2 | More interesting than mere publication count. |
| **Scimago Journal Rank (SJR)** | Citation PageRank of a journal divided by the number of articles pub-lished by the journal, in a 3 year citation period | Average per article PageRank based on Scopus citation data | Assigns different values to citations depending on the importance of the journals where they come from | Scopus is limited to the time period after 1996 for which citation analysis is available | 2 | 1 | Open access journals included in indicator |
| **Article influence score (AI)** | EigenFactor score divided by *i*-th entry in the normalized article vector | Measure of average per-article citation influence of the journal | Comparable to ISI JIF | Both EigenFactor and AI are redundant indicators as add little to easily understandable JIF, total citations and 5 year impact indicator (Chang et al. 2010) | 2 | 1 | Large disciplinary differences that persist in the Article Influence Score limit its utility for comparing journals across different fields (Arendt 2010) |
| **Normalised journal position (NJP)** (Bordons and Barrigon 1992; Costas et al 2012a) | Ordinal position of each journal in JCR category, ranked by JIF, divided by number of journals in that category. | Compare reputation of journals across fields | Allows for inter-field comparisons as it is a normalized indicator. | NJP is confounded by editorial decisions. All manuscripts have same rank position & the position is the result of successful publication decisions. | 2 | 2 | The citation counts of the published manuscripts determine the position of the journal (Bornmann et al. 2011) |
| **Diachronous IF** (Ingwersen et al. 2001) | A ratio calculation of citations from two or more citing years to documents issued in a fixed publication year | Reflects actual and development of impact over time of a set of papers. | Can be calculated for one-off publications, such as books containing contributions of different authors, or conference proceedings | Demands more resources than simply using impact factors from JCR, because it has to be based on manual collection of data. | 3 | 2 | Better represents the researcher in evaluation than SIF. |
| **CPP/FCSm** (Costas et al. 2010a) | Sum of citations divided by sum of world average | Individual performance compared to world citation average to publications of same document types, ages, and subfields. | Sum of citations before normalization makes indicator resistant to effect of highly cited papers in low-cited | Limited to same document type as world citation average is based on. | 3 | 3 | Calculation benefits older articles in highly cited fields (Moed 2005) |
| **CPP/JCSm** | Impact of individual's articles compared to average citation rate of individuals journal set. | Indicates if the individual's performance is above or below the average citation rate of the journal set. | Not affected by few publications that have a high/low citation count compared to world average. | Can be manipulated by publishing in averagely cited journals with a below average journal impact indicator (Moed 2005) | 3 | 2 | Citation rates are normalised as: the average citation rate of the researcher compared to average citation rate for field |
| **JCSm/FCSm** (Gaemers 2007. Costas et al. 2009;2010a;) | Journal citation score mean divided by field citation score mean | Relative impact level of the journals compared to their subfields | Normalised values are free from influences by distribution and document type effects. | The CPP/JCSm, CPP/FCSm and JCSm/FCSm indicators are not independent. The value of each one follows directly from the values of the other indices. | 3 | 3 | An unambiguous classification of articles in journals is impossible and different weighting schemes may lead to very different ratings in the evaluation |

# Appendix

| | | | | | | | |
|---|---|---|---|---|---|---|---|
| **C/FCSm** (van Leeuwen et al. 2003) | Total citation count divided by world mean citation rate of all publications in the same field (from same year of publication). | Applied impact score of each article/set of articles to the mean field average in which the researcher has published | Accepted as reliable measure for visibility in natural sciences. Highlights diversity of publication performance. | Unreliable due to non-para-digmatic nature of different fields, the heterogeneity of publication behaviours and insufficient coverage in citation databases. | 3 | 2 | Inadequate coverage in social and humanist sciences in citation indexes effects validity of indices. |
| **Prediction of article impact** (Levitt and Thelwall 2011) | Weighted sum of article citation and impact factor of the journal in which the article was published. | Predictor of long term citation | Aims to include new publications in analysis of an individual's research. | Indicator tested on only one subject category with a short publication window and may not apply to other subjects | 3 | 4 | Comparisons between the weighted sum indicator and the indicators from which it is derived (sum of citation and IF) need to be conducted with care. |
| **Co-authorship network analysis** (Yan and Ding 2011) | Weighted PageRank algorithm considering citation & co-authorship network topology | Individual author impact within related author community | Focuses on the random surfing aspect and develops it into citation ratios. | PR algorithm, only the top 10%-20% of overall authors in the co-authorship network can produce useful data. | 2 | 5 | Success of indicator is field dependent as rate of co-authorship varies |
| **Item oriented field normalized citation score average ($\overline{cf}$)** (Lundberg 2009) | Citations to individual publications divided by world average of citations to publications of the same type, year and subject area | Item orientated field normalised citation score. | Normalisation is on the level of individual publication giving each publication equal weight in the final field score value. Accounts for the prevailing skewness of citation distributions | Value of field normalised citation score can be unproportionately affected by highly cited publications in a moderately cited field. | 3 | 4 | More appropriate for some document types than others; there are differences in average availability of citation data, citation rates, and document types used in research. |
| **%HCP** (Costas et al. 2010a) | publications cited above the 80-percentile in their respective research areas | Indicates papers among the 20% most cited in research area, i.e relative impact | Indicator of excellence understood as citation count reflect the extent to which an academic's work affects the work of his/her peers | Indicates only one facet of excellence and no reflection of the impact of the work on society | 3 | 4 | Difficult to maintain high values of relative impact with increasing rates of production (Costas et al. 2009) |

* Col. = data collection, Cal. = calculation

# Appendix

## Effect of output as citations, category 3a

| Indicator | Definition | Designed to indicate | Individual | | Complexity | | Comments |
|---|---|---|---|---|---|---|---|
| | | | Advantages | Limitations | Col* | Cal* | |
| **Nnc** | Number not cited | The sum of uncited papers | can contextualise the number of papers not cited to academic age or used to explain performance on other such as CPP | Does not indicate lack of citation means lack of quality or usefulness | 1 | 1 | Illustrates the types of publications although important that do not receive citations, i.e. technical reports, guidelines ets. |
| **Database dependent counting i.a.** (Scimago total cites, WoS, Scopus, Google Scholar) | Number of citations recorded in the selected database | The number of citations is dependent on the database used to collect the citation information | Indicates how coverage of researcher in database can effect calculation of bibliometric indicators and performance of researcher | Many of the more sophisticated indicators and field benchmarks are reliant on WoS and as such cannot be compared with data from other sources | 2 | 1 | Scope, validity, reliability and cost of the citation collection is dependent on choice of citation index. |
| **i10-index** (Google Scholar Metrics) | The number of publications with at least 10 citations. | Number of substantial papers. The worth of "10" citations is highly field dependent. | Includes a broad source/publication base; simple and straightforward to calculate and Google Scholar My Citations is free and easy to use. | Very simple and easy to modify citation profiles in Google Citations and fraudulently affect Google Metrics (Delgado López-Cózar 2014) | 2 | 1 | i10 is very sensitive. The publication and citation data needs careful verification as Google Scholar is criticized for misattributing documents and cited references (Jascó 2011) |
| **C** | Number of citations recorded in CI†, minus self-citations | Recognised benchmark for analyses. Indication of usage by stakeholders for whole period of analysis | Reflects social side of research and the cumulative development of knowledge in CI processed publications | Quality and timeliness of citation not considered; | 2 | 2 | Does not account for older articles being more cited and variation of citation rates between document types and fields. |
| **Sc** | Sum of self-citations | Indication of self-use for whole period of analysis | Illustrates how work builds on previous findings. Advertises the work and the author | Unclear what a self-citation is: cites of oneself, a co-author or institutional colleague. | 2 | 2 | Self-citation is highly variable among individuals and its contribution highly variable., Self-citations are not dismissible when calculating citation statistics |
| **C + sc** | Count of all citations to all or selected output, including self-citations | Indication of all usage for whole period of analysis | Reflects social side of research and the cumulative development of knowledge | Quality and timeliness of citation not considered | 2 | 1 | Self-citations affect the reliability & validity of the measure on small amounts of data in assessments ( Glänzel, et al. 2006; Costas and Bordons 2007) |
| **Fractional citation count** (Egghe 2008) | Gives an author of an m-authored paper only credit of c/m if the paper received c citations | Designed to remove the dependence of co-authorship (Egghe, 2008) | Gives less weight to collaborative works and leads to proper normalization of indicators and fairer comparisons | Regards credit as a single unit that can be distributed evenly, making share dependent on number of authors. | 2 | 2 | Comparison to field norm unwise as citations to the publications may not be representative of the field but biased towards the highly or poorly cited. |

# Appendix

| | | | | | | | |
|---|---|---|---|---|---|---|---|
| **C-sc** | Citation count, self-citations removed | Measure of usage for whole period of analysis | Reflects social side of research and the cumulative development of knowledge | Quality and timeliness of citation not considered; Unclear what to exclude: cites of oneself, a co-author or institutional colleague. | 2 | 2 | Does not account for older articles being more cited and variation of citation rates between document types and fields. |
| **MaxC** | Paper with the highest number of citations | The most significant paper | Simple indicator of the most important and influential research. | The most significant paper is not necessarily the paper with the most citations | 2 | 2 | The most highly cited papers can have the largest number of authors, tend to be longer than the average article and have more references. |
| **Citations in patents** (Okubu 1997) | Count and source assessment of citations in patents | Impact on or use in new innovations | Depicts state of a given art, newness and significance of innovation; length of time between publication of paper and patent application. | Cites might be legally or competitively motivated and not of innovative or scientific nature. Indicates impact of technology rather than science | 4 | 1 | Requires access to specialized database and cooperation of several specialists to verify results (Quomiam et al. 1993) |
| **Knowledge use** (Mostert 2010) | Count of use of output in schoolbooks, curriculum, protocols, guidelines, policies and in new products | Impact on learning in stakeholders environment. | Analysis of citations and references in guidelines, policies, protocols to indicate links (use) with stakeholders. | Has to be adjusted to the mission and objectives of the scientist and department/discipline | 5 | 1 | Focuses on research group level |

* Col. = data collection, Cal. = calculation
†CI =Web of Science (CI) versions of the Science Citation Index, the Social Science Citation Index, Arts and Humanities Citation Index

# Appendix

## Effect as citations normalized to publications and field, 3b

| Indicator | Definition | Designed to indicate | Individual | | Complexity | | Comments |
|---|---|---|---|---|---|---|---|
| | | | Advantages | Limitations | Col* | Cal* | |
| **Tool to measure societal relevance** (Niederkrotent-haler et al. 2011) | Questionnaire used as the (self-assessment) application form and the assessment form for the reviewer | Aims at evaluating the the level of the effect of the publication, or at the level of its original aim | Accounts for knowledge gain, application &increase in awareness; efforts to translate research results into societal action; identification of stakeholders and interaction with them. | Only developed and evaluated in a focus group in the biomedical sciences | 1 | 1 | Tool requires further development, specification and validation. |
| **Number of significant papers** | Papers with >y citations, | Gives idea of broad and sustained impact | y can be adjusted for seniority, field norm and publication types | Subjective. | 2 | 1 | Can randomly favour or disfavour individuals |
| **Field top % citation reference value** | Quota between count of publications in group, as above, and those with citations above n%. | World share of publications above citation threshold for n% most cited for same age, type and field | Percentiles can prevent a single, highly cited publication receiving an excessively heavy weighting | The degree to which top n% publications are over/under-represented differs across fields and over time (Waltman and Schreiber 2012) | 3 | 3 | Accuracy of inter-field and inter-temporal comparisons decreases with level of representation. |
| **E(Ptop)** | Expected number of highly cited papers among the top 20, 10, 5, 1% in the subfield/world | Reference value: expected number of highly cited papers based on the number of papers published by the research unit. | Reflects deviations from the 80th, 90th, 95th, 98th, 99th percentile if tied values occur due to the discrete nature of the impact distribution. | Only Includes documents that have been cited at least once and is interpreted as normalised citations per cited paper not citations per paper | 3 | 3 | Expected scores are based on large data sets, their 'random' error is much smaller than that of the value CPP. |
| **A/E(Ptop)** | The ratio of the actual and expected presence in the top of the citation distribution. | Relative contribution to the top 20, 10, 5, 2 or 1% most frequently cited publications in the world relative to year, field and document type. | Indicates share of top impact publication. | Does not account for time delays between publication and citations | 3 | 3 | Can reveal if a high normalized score is due to a few highly cited papers or a general high level of citations. |
| **Index of Quality and Productivity** (Antonakis and Lalive 2008) | Ratio actual citations to estimated citations and total papers (corrected for subject) | Quality reference value; judges the global number of citations a scholar's work would receive if it were of average quality in its field. | Corrects citation count for scholarly productivity, author's academic age, and field-specific citation habits with reference to estimated citation rate. | Tested in natural sciences, medicine and psychology and dependent on WOS field specific journal impact factors. | 3 | 3 | Correlates better with expert ratings of greatness than h index. Allows comparison as brings papers in low cited fields on same scale as papers in highly cited fields. |
| **Ptop** | Publications are grouped by type, age and subject, then ranked by citations. | Identify if publications are among the top 20, 10, 5, 1% most frequently cited papers in subject/subfield/world in a given publication year. | Indicates if publications are cited well but fail to produce high impact or if researcher contributes to high impact publications but also has a pool of less well cited work. | Percentiles are most suitable for normalisation of citation counts in terms of subject, document type and publication year (Bornmann and Werner 2012) | 3 | 3 | Unlike mean based indicators, percentiles are not affected by skewed distribution |
| **Scientific proximity** (Okubu 1997) | Relative number of citations of papers in patents applied for in specific sector | Intensity of an industrial or technological activity | Interaction between science and technology | Credibility of any utilisation of such data for analytical and statistical purposes. | 5 | 2 | Patents serve a legal purpose, and authors demonstrate their technological links and conceal the essentials of their content |

* Col. = data collection, Cal. = calculation

# Appendix

## Effect as citations normalized to publications in portfolio, 3c

| Indicator | Definition | Designed to indicate | Individual | | Complexity | | Comments |
|---|---|---|---|---|---|---|---|
| | | | Advantages | Limitations | Col* | Cal* | |
| **Age of citations** | Identifies how old citations are. | If a large citation count is due to articles written a long time ago and no longer cited OR articles that continue to be cited. | Accounts for differences between delayed citations and sleeping beauties, and inter-field differences (van Raan 2004) | Observed age of citations may not conform with theoretical distributions as the measure cannot cope with singularities from usage of literature on a micro level (De Bellis 2009) | 3 | 1 | Usage and validity are not directly related and might merely reflect the availability of documents. |
| **%Pnc** | Number of non-cited publications divided by total number publications in same time period | Share of publications never cited after certain time period, excluding self-citations | Benchmark value: cited and non-cited publications reflect their underlying relevance for technological developments | Publications can be greatly used and of great influence, but never cited (MacRoberts and MacRoberts 2010) | 3 | 1 | Endorses utilitarian approach: "useful" (i.e. cited within 5 years post-publication), and "not useful/useless" (not cited within 5 years) |
| **% SELFCIT** | Number of self-citations divided by total citations | Share of citations to own publications | Reflects readership of work outside of author and group. | Unclear what to exclude: cites of oneself, a co-author or institutional colleague | 3 | 2 | Identifies unwarranted self-promotion |
| **%nnc** | Percent not cited | Share of uncited publications | Using a non-citation rate is advantageous as it creates a clear distinction between how citation analysis is used to determine the quality of researcher | Authors cite only a fraction of their influences, many citations go to secondary sources, and that informal level of communication is not captured | 1 | 2 | Can be used to compare authors and the percentage of their work that has not been cited to the present date |
| **CPP** | Sum of citations divided by number of publications. | Trend of how cites evolve over time | Enables comparisons of scientists of different ages and different type of publications | Tells nothing of the timeliness, origin or quality of the cite (positive or negative) | 2 | 2 | Citations can be hard to find, reward low productivity & penalize high productivity (Haslam and Laham 2009) . |
| **MedianCPP** (Grothkopf and Lagerstrom 2011) | The median number of citations per paper | Trend of how cites evolve over time, adjusting for skewness (variation in spread of citations) | Less sensitive to high or non-cited publications than CPP. | Possible information redundancy as significant correlations between CPP, median CPP, h and g-index have been found (Calver and Bryant 2008) | 2 | 3 | Suited for the comparisons of scientists who have been active for different number of years |

* Col. = data collection, Cal. = calculation

# Appendix

## Indicators that rank the portfolio: h dependent, category 4a

*All indices require verified publication data and data from one or more citation databases*

| Indicator | Definition | Designed to indicate | Individual | | Complexity | | Comments |
|---|---|---|---|---|---|---|---|
| | | | Advantages | Limitations | Col* | Cal* | |
| **h-index** (Hirsch 2005) | Publications ranked in descending order by the times cited. H is the number of papers (N) in the list that have N or more citations. | Cumulative achievement | H is a simple but rough measurement of quality of work, when compared to JIF, citation & publication count (Alonso et al. 2009) | Once a paper is in H-core, the number of citations it receives is disregarded. Loss of citation information means comparisons based on the h-index can be misleading (Schreiber et al. 2012) | 2 | 2 | Arbitrary cut off value for including or excluding publications from productive h-core. |
| **m-index** (Bornmann et al. 2008) | Median number of citations received by papers in the h-core | Impact of papers in the h-core | To account for skewed distribution of citations, the median and not the arithmetic average is used to measure a central tendency. | Although median may be a better measure of central tendency it can be chronologically instable. | 2 | 2 | Reduces impact of heavily cited papers. |
| **e-index** (Zhang 2009) | E is the number of excess citations (more-than-h citations received by each paper in the h core) | Complements the h-index for the ignored excess citations | The combination *h,e* provides complete citation information. | E value can only be calculated if h is given. | 2 | 2 | Complements h especially for evaluating highly cited scientists or for precisely comparing the scientific output of a group of scientists having an identical h-index. |
| **Hmx-index** (Sanderson 2008) | Rank academics by their maximum *h* (hmx) measured across WOS, Scopus and GS. | Ranking of the academics using all citation databases together. | Accounts for missing citations, lack of correlation between databases and disparities in h across databases. | Assumes that the differences in h across the databases are due to false negative errors and that these were negligible. | 2 | 2 | Although hmx provides a better estimate of h than any single database, a close examination of the overlaps of citations and publications between the databases will provide a better estimate. |
| **Hg-index** (Alonso et al. 2009b) | Geometric mean of a scientist's h- and g-indices, i.e. $hg=\sqrt{h.g}$ | Greater granularity in comparison between researchers with similar h- and g- indices. | Accounts for influence of a big successful paper on g-index to achieve balance between the impact of the majority of the best papers of the author and very highly cited ones. | Combining H and G does not improve discriminatory power, hg has no direct meaning in terms of papers and citations of a scientist and can lead to hasty judgements (Franceschini and Maisano 2011) | 2 | 3 | Simple to compute once the h- and g-indices have been obtained. |
| **H(2) index** (Kosmulski 2006) | The highest natural number such that the scientist's H(2) most cited papers received each at least H(2)2 citations. | Weights most productive papers but requires a much higher level of citation attraction to be included in index. | Precision/homograph problem reduced as only a small subset of the researcher's papers used to calculate H(2) index (Bornmann et al. 2008; Jin et al. 2007) | Difficult to discriminate between scientists having different number of publications with quite different citation rates for relatively high H(2) indices | 2 | 3 | Suffers from same inconsistency problems as h. (Waltman and van Eck 2011) |

# Appendix

| | | | | | | | |
|---|---|---|---|---|---|---|---|
| **A-index** (Jin 2006; Rousseau 2006) | Average number of citations in h-core thus requires first the determination of h. | Describes magnitude of each researcher's hits, where a large a-index implies that some papers have received a large number of citations compared to the rest (Schreiber et al. 2012) | a-index can increase even if h-index remains the same as citation counts increase (Alonso et al. 2009) | a is h-dependent, has information redundancy with h, and when used together with h masks the real differences in excess citations of different researchers (Schreiber et al. 2012) | 2 | 3 | A-index involves division by h and punishes researchers with high h-index (Jin et al. 2007) ; sensitive to highly cited papers (Rousseau 2006) |
| **R-index** (Jin et al. 2007) | Square root of the h and A index | Citation intensity and improves sensitivity and differentiability of A index | Adjusts for punishing the researcher with a high h index; | As above. R-index involves division by h and punishes researchers with high h-index; (Jin et al. 2007); | 2 | 3 | Supplement to h. Easier to calculate than g index, but not as elegant. |
| **h-index** (Miller 2006) | Square root of half the total number of citations to all publications | Comprehensive measure of the overall structure of citations to papers | Includes papers h ignores ie. most highly cited articles and the body of articles with moderate citations | Difficult to establish the total citation count with high precision (Schreiber 2010) | 2 | 3 | Is only roughly proportional to h. |
| **$Q^2$ –index** (Cabrerizoa et al 2012) | $Q^2$ is the geometric mean of h-index and the median number of citations received by papers in the h-core | Relates two different dimensions in a researcher's productive core: the number and impact of papers | Combines robustness of h-index' measurement of papers in core with m-index correction of the distribution of citations to papers. | h- and m-indices have to be obtained before calculation of $q^2$ | 2 | 3 | Geometric mean is not influenced by extremely higher values, and obtains a value which fuses the information provided by the aggregated values in a balanced way. |
| **H per decade (Hpd-index)** (Kosmulski 2009) | Hpd is highest number of papers that have at least hpd citations per decade each and other papers have less than hpd + 1 citations per decade each. | Compare the scientific output of scientists in different ages. Seniority-independent Hirsch-type index. | In contrast with h-index, which steadily increases in time, hpd of a mature scientist is nearly constant over many years, and hpd of an inactive scientist slowly declines. | Hpd uses scaling factor of 10 to improve granularity between researchers is as an arbitrary number, which randomly favors or disfavors individuals. | 2 | 4 | hpd can be further modified for multi-authored papers where the individual cites per year of each paper is divided by the number of co-authors to produce the contribution of single co-author. |
| **Citation-weighted h-index (hw)** (Egghe and Rousseau 2008) | Hw is the square root of the total weighted citations (Sw) received by the highest number of articles that received Sw/h or more citations | Weighted ranking to the citations, accounting for the overall number of h-core citations as well as the distribution of the citations in the h-core. | Improves sensitivity to the number of citations in h-core | Does not use h-table in calculation and is therefore not an acceptable h-type measure | 2 | 4 | Hw can be misleading and a contradiction of h (Maabreh and Alsmadi, 2012) |
| **hα** (Eck and Waltman 2008) | The value of hα is equal to N papers with at least α· hα citations each and the other n- Hα papers have fewer than ≤ α· hα citations each. | Cumulative achievement, advantageous for selective scientists. | Greater granularity in comparing scientists with same h is possible; α can be set to the practices in a specific field, allowing for fairer comparison between fields. | No agreement on the value of parameter α. The appropriate choice of α requires more study and is field dependent. Sensitivity of hα to α needs investigating. | 2 | 4 | Small α: ranks scientists based on number of papers with at least one citation (quantity measure: advantageous for scientist who publish a lot but are not very highly cited) Large α: measures number of citations of most cited paper (quality). |

# Appendix

| | | | | | | | |
|---|---|---|---|---|---|---|---|
| **b-index** (Brown 2009) | B is the integer value of the author's external citation rate (non-self-citations) to the power three quarters, multiplied by their h-index | The effect of self-citations on the h-index and identify the number of papers in the publication set that belong to the top n% of papers in a field | Cut-off value for including or excluding publications in productive core is determined using a field-specific reference standard for scientific excellence (Bornmann et al. 2007) | Assumes that relative self-citation rate is constant across an author's publications | 2 | 4 | The b index depends on the year in which it is determined, the period under consideration and the used database |
| **Tapered h-index (hT)** (Anderson et al. 2008) | Using a Ferrers graph, the h-index is calculated as equal to the length of the side of the Durfee square assigning no credit to all points that fall outside. | Production and impact index that takes all citations into account, yet the contribution of the h-core is not changed. | Evaluates the complete production of the researcher, all citations giving to each of them a value equal to the inverse of the increment that is supposed to increase the h-index one unit. | Difficult to implement because of the computations needed to obtain the measure and the difficulty in obtaining accurate data from bibliographic databases (Alonso et al 2009). | 2 | 5 | Shows smooth increase in citations, not irregular jumps as in h-index. Conceptually complex (Anderson et al 2008). |
| **Rational h-indices hrat Index** (Ruane and Tol 2008) | hrat=(h+1) – $\frac{nc}{2.h+1}$ h is h index, nc is number of citations that are needed to make a h-index of h+1 and 2. | Indicates the distance to a higher h-index by interpolating between h and h+1. h+1 is the maximum amount of cites that could be needed to increment the h index one unit (Alonso et al 2009). | Increases in smaller steps than h-index providing greater distinction in ranking of individuals | The relative influence of the interpolation will be stronger for smaller values of the indices therefore utilize the generalized indices when comparing many data sets with very small values of h. | 2 | 5 | Interpolated indices have the advantage that one does not have to wait so long to see one’s index growing. |

* Col. = data collection, Cal. = calculation

# Appendix

## Indicators that rank the portfolio: h independent, category 4b

*All indices require verified publication data and data from one or more citation databases*

| Indicator | Definition | Designed to indicate | Individual | | Complexity | | Comments |
|---|---|---|---|---|---|---|---|
| | | | Advantages | Limitations | Col* | Cal* | |
| **w-index** (Wu 2008) | w is the highest number of papers have at least 10w citations each | The integrated impact of a researcher's excellent papers. | More accurately reflects the influence of a scientist's top papers | Tendency to describe quantity of the productive core | 2 | 2 | w-index of 1 or 2 is someone who has learned the rudiments of a subject; 3 or 4 is someone who mastered the art of scientific activity, while "outstanding individuals" have a w-index of 10. |
| **g-index** (Egghe 2006) | Publications ranked in descending order by times cited. G is highest number g of papers that together received g2 or more citations | The distinction between and order of scientists (Egghe, 2006; Harzing, 2008) | Corrects h by weighting highly cited papers to make subsequent citations to highly cited papers count in calculation of the index. | Can be disproportionate to average publication rate. The G-index of a scientist with one big hit paper and a mediocre core of papers could grow in a lot comparison with scientists with a higher average of citations | 2 | 3 | Ignores the distribution of citations as based on arithmetic average. (Costas and Bordons 2007; Alonso et al. 2009) |
| **f-index** (Tol 2009) | Fractional counting and ranking scheme of papers:cites, where the average is calculated as the harmonic mean | Attempts to give weight/value to citations. Highest number of articles that received *f* or more citations on average. | An additional citation to a not-so-often cited paper counts more than an additional citation to an often-cited paper. | Both f & t indices are maximum if every paper is cited the same number of times, but the f-index deviates much faster from this maximum than the t-index. | 2 | 3 | More discriminatory power than the h- and g-indices. Because of the non-linearity of the harmonic mean, the f-index is more sensitive to small differences between researchers |
| **t-index** (Tol 2009) | Fractional counting and ranking scheme of papers:cites, where the average is calculated as the geometric mean | Attempts to give weight/value to citations. Highest number of articles that received t or more citations on average | Using geometric mean doesn't place much weight on the distribution of citations. | Sensitivity to small differences between researchers is stronger with harmonic mean (f-index) than geometric mean. | 2 | 3 | It is not sufficient to determine the function and value of citations using indices; their cognitive background should also be taken into consideration. |
| **π-index** (Vinkler 2009) | π is one hundredth of the number of citations received by the top square root of the total number of papers ranked by decreasing number of citations. | Production and impact of scientist | Allows for comparative assessment of scientists active in similar subject fields. Sensitive to citedness of top papers and thus indicates impact of information on research. | Value depends on citation rate of papers in the elite set (top cited papers); the elite set is scaled by an arbitrary prefactor (Schreiber 2010). | 2 | 4 | Can be calculated on a small number of papers. Unique index because it is defined in terms of the summed number of citations rather than the square root of the sum or the average (Schreiber 2010). |

# Appendix

| | | | | | | | |
|---|---|---|---|---|---|---|---|
| **Gα** (Eck and Waltman 2008) | gα is the highest rank such that the first gα papers have, together, at least citations. | Based on same ideas as g-index, but allows for fractional papers and citations to measure performance at a more precise level. | gα-index puts more weight on the quality aspect of scientific performance than the hα-index. | No agreement on the value of parameter. The appropriate choice of Gα requires more study and is field dependent. | 2 | 4 | Empirical research is needed to find out whether in practical applications the gα index provides better results than g-index |
| **Rational g-index grat,** (Schreiber 2008a; Tol 2008) | Interpolates between g and g+1 based as above on the piecewise linearly interpolated citation curve. | Indicates the distance to a higher g-index | It is not a complementary index requiring first the determination of g, but rather follows from a self-consistent definition (Schreiber 2010) . | Limits as for hrat. | 2 | 5 | As every citation increases interpolated g, the index is sensitive to self-citations (Schreiber 2008a) |

* Col. = data collection, Cal. = calculation

# Appendix

## Indicators that rank the portfolio: h adjusted to field, category 4c

*All indices require verified publication data and data from one or more citation databases*

| Indicator | Definition | Designed to indicate | Individual | | Complexity | | Comments |
|---|---|---|---|---|---|---|---|
| | | | Advantages | Limitations | Col* | Cal* | |
| **n-index** (Namazi and Fallahzadeh 2010) | Researcher's h-index divided by the highest h-index of the journals of his/her major field of study | Enables comparison of researchers working in different fields: | Can surmount the problem of unequal citations in different fields | Still awaiting validation. | 2 | 2 | Calculation based on Scopus definition of h and SCImago Journal and Country Rank website for journal information |
| **Normalized h-index** (Sidiropoulos et al. 2007) | hn =h/Np, if h of its Np articles have received at least h citations each, and the rest (Np−h) articles received no more than h citations. | Normalizes h to compare scientists achievement based across fields | Accounts for the fact that scientists have different publication and citation habits in different fields. | The normalized h-index can only be used in parallel to h-index and rewards less productive but highly cited authors | 2 | 3 | Using this parameter to judge someone still at the beginning of their career, with few publications, is prone to give paradoxical results. |
| **h-index sequences and matrices** (Liang 2006) | Calculates h-sequence by continually changing the time spans of the data. Constructs h-matrix based on a group of correlative h-sequences. | Singles out significant variations in individual scientists citation patterns across different research domains | Makes scientists of different scientific age comparable. | Difficult to determine the correct publication/citation window in construction of the matrix | 2 | 4 | Only tested on 11 well established physicists. |
| **Generalized h-index hf** (Radicchi et al. 2008) | Citations of each article normalized by average number of citations per article in the subject category of the article under observation | Allows comparison to peers by correcting individual articles' citation rates for field variation | Suitable for comparing scientists in different fields as rescales field variations and factors out bias of different publication rates | Scales number of citations and rank of papers by constants dependent on discipline, however constants are not available for all fields. | 3 | 4 | Calculation is not easy making it a nominal index and not a pragmatic one (Namazi and Fallahzadeh 2010) |
| **x-index** (Claro & Costa 2011) | x is a researcher's absolute score divided by a reference score | Indication of research level. Describes quantity and quality of the productive core and allows for comparison with peers. | Accounts for multi-and interdisciplinary research by using the journals the researcher publishes in as reference and not field classification | x is based on (5 year) Impact Factor which has well-documented limitations; x is also vulnerable to scale issues | 3 | 4 | Using a measure based on citation counts would permit a more meaningful assessment of scientific quality |

*Col. = data collection, Cal. = calculation

# Appendix

## Indicators that rank the portfolio: h corrected for co-authorship, category 4d

*All indices require verified publication data and data from one or more citation databases*

| Indicator | Definition | Designed to indicate | Individual | | Complexity | | Comments |
|---|---|---|---|---|---|---|---|
| | | | Advantages | Limitations | Col* | Cal* | |
| **Alternative H index** (Batista et al. 2006) | Alternative h is h-index divided by mean number of authors in the h publications | Indicates the number of papers a researcher would have written along his/her career if worked alone. | Rewards scientists whose papers are entirely produced by themselves from the authors that work groups that publish a larger amount of papers. Accounts for differences in co-authorship patterns, disciplinary differences and self-citations (Schreiber 2008a) | Mean is sensitive to extreme values and could penalize authors with papers with a large number of authors. Might decrease when a paper with many authors advances into the h-core by attracting additional citations and reduces size of the h-core. | 2 | 2 | Valid quantification of output across disciplines allowing for comparison. |
| **POP variation individual H-index** (Harzing 2008) | Divides number of citations by number of authors for that paper, then calculates the h-index of the normalised citation counts | Accounts for co-authorship effects | Gives an approximation of the per-author impact, which is what the original h-index set out to provide. | Normalisation by mean number of authors of publications in the h-core leads to reduction of the index. This is a fractionalised count of citations and publications (Schreiber 2008a) | 2 | 3 | (Egghe 2008) also considered multiple authors by computing g and h indices using a fractional crediting system. |
| **Pure h-index (Hp)** (Wan et al. 2007) | Hp is the square root of h divided by normalised number of authors and credit to their relative rank on the by-line of the h-core articles | Corrects individual h-scores for number of co-authors | Reduces effect of collaboration in multi-authored, highly cited paper. | Results vary dependent on method of distributing credit to authors- fractional count, arithmetic to determine h, | 2 | 4 | More refined approach is pure R-index. Takes the number of collaborators, possibly the rank in the byline and the actual number of citations into account. |
| **$H_m$-index** (Schreiber 2008b) | Uses inverse number of authors to yield a reduced or effective rank. Hm is the reduced number of papers that have been cited hm or more times | Softens influence of authors in multi-authored papers | Does not push articles out of the h-core; each paper is fully counted allowing for a straightforward aggregation of data sets. | Precision problem is enhanced, as additional papers enter into the hm-core. | 2 | 4 | Uses fractional paper counts instead of reduced citation counts |
| **Adapted pure H-index ($h_{ap}$)** (Chai et al. 2008) | H is interpolated rank value between papers (fractionally counted) and citations (counted as square root of equivalent number of authors). | Finer granularity of individual h-scores for number of co-authors by using a new h-core. | Alters h-core to be less biased than Hp with respect to authors with many multi-authored papers | Precision an issues and difficult to calculate. | 2 | 5 | Lead to a more moderate correction of authorship than $h_i$ as divides citation count by the square root of author count rather than full author count (Rosenberg 2011) |

* Col. = data collection, Cal. = calculation

# Appendix

## Indicators of impact over time: normalized to portfolio, category 5a

*All indices require verified publication data and data from one or more citation databases.*

| Indicator | Description | Designed to indicate | Individual | | Complexity | | Comments |
|---|---|---|---|---|---|---|---|
| | | | Potentials | Limitations | Col* | Cal* | |
| **Age-weighted citation rate (AWCR, AW & per-author AWCR)** (Harzing 2012b) | Age-weighted citation rate, is the number of citations to a given paper divided by the age of that paper | AWCR measures the number of citations to an entire body of work, adjusted for the age of each individual paper | Using the sum over all papers instead, represents the impact of the total body of work allowing younger, less cited papers to contribute to the AWCR | Field norm has to be decided to account for field characteristics such as expected age of citations, “sleeping beauties”, and delayed recognition. | 2 | 3 | The AW-index is defined as the square root of the AWCR. It approximates the h-index if the mean citation rate remains constant over the years. The per-author age-weighted citation rate is similar to the plain AWCR, but is normalized to the number of authors for each paper. |
| **AR-index** (Jin, Liang, Rousseau, & Egghe 2007) | AR is the square root of the sum of the average number of citations per year of articles included in the h-core. | Accounts for citation intensity and the age of publications in the core. | AR is necessary to evaluate performance changes. | Divides the received citation counts by the raw age of the publication. Thus the decay of a publication is very steep and insensitive to disciplinary differences. (Järvelin and Person, 2008) | 2 | 2 | AR index increases and decreases over time (Alonso et al 2009); Complements h. Jin et al do not consider AR convincing as a ranking metric in research evaluation. |
| **M-quotient** (Hirsch 2005) | M is h-index divided by years since first publication | H type index, accounting for length of scientific career | Allows for comparisons between academics with different lengths of academic careers, as h is approximately proportional to career length. | m stabilizes later in career; small changes in h can lead to large changes in m; first paper not always an appropriate starting point (Harzing 2008) | 2 | 2 | Although the m-quotient adds time as a weighting factor, it does not cater to the major disadvantages of the h-index including quality of publication and quality of citation |
| **Mg-quotient** | M is g-index divided by years since first publication | G type index, accounting for length of scientific career | Allows for comparisons between academics with different lengths of academic careers, as h is approximately proportional to career length. | First publication is not necessarily the appropriate estimate of the start of the researcher’s career | 2 | 3 | mg and m discriminate against part time researchers/career interruptions |
| **Price index – PI** (Price 1970) | PI = (n1/n2)*100 where n1, is the num-ber of cited references with a relative age of less than 5 years, n2 is the total number of references. | Percentage references to documents, not older than 5 years, at the time of publication of the citing sources | Accounts for the differing levels of immediacy characteristic of the structurally diverse modes of knowledge production occurring in the different sciences | Does not reflect the age structure in slowly ageing literature (De Bellis, 2009) | 3 | 2 | In the calculation of PI it is unclear whether the year of publication, is year zero or year one. Moreover, it is unclear whether or not this year is included. (Egghe & Rousseau 1995) |

# Appendix

| | | | | | | | |
|---|---|---|---|---|---|---|---|
| **Citation age c(t)** (Egghe & Rousseau 2000) | c(t) is the difference between the date of publication of a researcher's work and the age of citations referring to it. | The age of citations referring to a researcher's work. | The entire distribution of the citation ages of a set of citing publications provides insight into the level of obsolescence or sustainability. | Possibility of measuring aging in a meaningful way is questionable by means of citation counting as this doesn't account for role of literature growth, availability of literature and disciplinary variety | 3 | 3 | Usage and validity are not necessarily related |
| **Contemporary h-index $h^c$** (Sidiropoulos et al. 2007) | An article is assigned a decaying weight depending on its age | Currency of articles in h-core. | Accounts for active versus inactive researchers | The weighting is parametrized gamma=4 and delta=1, making this metric identical to hpd, except measured on a four year cycle rather than a decade. (Rosenberg, 2011) | 2 | 4 | An old article gradually loses its "value", even if it still gets citations thus newer articles are prioritized in the count. |
| **Trend H index $h^t$** (Sidiropoulos et al 2007) | Each citation of an article is assigned an exponentially decaying weight, which is expressed as a function of the "age" of the citation. | Age of article and age of citation. | Identifies pioneering articles that set out new line of research and still cited frequently. | The weighting is parametrized and for gamma = 1 and delta = 0, this metric is the same as the h-index. | 3 | 4 | Estimates impact of researchers work in a particular time instance i.e. whether articles still get citations by looking at the age of the cites. |
| **Dynamic H-type index** (Rousseau & Ye 2008) | Built on 3 time dependent elements: R(T)· vh(T) where R(T) is the R-index computed at time T and vh is the h-velocity | Accounts for the size and contents of the h-core, the number of citations received and the h-velocity. | Detects situations where two scientists have the same h index and the same number of citations in the h core but that one has no change in his h index while another scientist's h index is on the rise. | H dependent. To define vh it is better to find a fitting for hrat(t) - and not for h(t)- as this function is more similar to a continuous function than the standard h-index. | 3 | 4 | For evaluation purposes self-citations should be removed (Alonso et al 2009). |
| **Discounted Cumulated Impact (DCI)** ( Järvelin and Person 2008; Ahlgrena & Järvelin 2010) | Sum of weighted count of citations over time to a set of documents divided by the logarithm of the impact in past time intervals | Devalues old citations in a smooth and parameterizable way and weighs the citations by the citation weight of the citing publication to indicate currency of a set of publications. | Gives more weight to highly cited publications as these are assumed to be quality works. | Difference caused by weighting: some authors gain impact while some others lose. | 3 | 5 | Rewards an author for receiving new citations even if the publication is old. |

* Col. = data collection, Cal. = calculation

# Appendix

## Indicators of impact over time: normalized to field, category 5b

*All indices require verified publication data and data from one or more citation databases.*

| Indicator | Description | Designed to indicate | Individual | | Complexity | | Comments |
|---|---|---|---|---|---|---|---|
| | | | Potentials | Limitations | Col* | Cal* | |
| **Classification of durability** (Costas 2010a; 2010b; 2011) | Percentile distribution of citations that a document receives each year, accounting for all document types and research fields. | Durability of scientific literature on distribution of citations over time among different fields | Aids study of individuals from general perspective using composite indicators. Discriminates between normal, flash in the pan and delayed publications. | Minimum 5 yr citation history threshold for reliable results and empirically investigated in WOS using journal subject categories. | 2 | 4 | Can be applied to large sets of documents or documents published in different years; Documents can be classified in more than one field and can be updated yearly/monthly |
| **Aging rate a(t)** (Egghe & Rousseau 2000) | a(t) is the difference between ct and c(t+1) | Aging rate of a publication. | For individual documents stochastic models are preferable as they allow for translation of diverse factors influencing aging into parameters that can be estimated from empirical data with a specified margin of error | A corrective factor is required if citation rates are to be adjusted for changes in the size of citing population and discipline (De Bellis 2009; Dubos 2011) | 3 | 4 | There are many models to study aging, the simplest is study of the exponential decay of the distribution of citations to a set of documents |
| **Age and productivity** (Costas et al. 2010a) | Mean number of documents by age and CPP (3 yr citation window) in 4 year age brackets, adjusted to field. | Effects of academic age on productivity and impact. | Identifies the age at which scientists produce their best research and the extent of the decline in their production | Mean impact declines with age regardless of quality of researcher's body of work. | 4 | 4 | If used independently, fosters practice of quantity over quality. Difficult to maintain high values of impact with increasing rates of production. |

* Col. = data collection, Cal. = calculation

# Appendix

# Appendix

# Appendix

# Appendix

# Appendix

# Appendix

## Appendix

# Appendix

# Appendix